\documentclass[fleqn,usenatbib]{mnras}

\usepackage{newtxtext,newtxmath}

\usepackage[T1]{fontenc}

\DeclareRobustCommand{\VAN}[3]{#2}
\let\VANthebibliography\thebibliography
\def\thebibliography{\DeclareRobustCommand{\VAN}[3]{##3}\VANthebibliography}

\usepackage{graphicx}	
\usepackage{amsmath,commath,mathtools}
\usepackage{float}
\usepackage{booktabs}
\usepackage{ulem}
\usepackage{color}
\usepackage{xcolor,colortbl}

\newcommand{\Ngrid}{N_{\text{grid}}}
\newcommand{\Nsol}{N_{\text{sol}}}
\newcommand{\Ntot}{N_{\text{tot}}}
\newcommand{\Ncore}{N_{\text{core}}}
\newcommand{\pol}{\ensuremath{p}}
\newcommand{\xcoord}{\ensuremath{ {\boldsymbol x} }}
\newcommand{\kvec}{\ensuremath{ {\boldsymbol k} }}
\newcommand{\Psivec}{\ensuremath{ {\boldsymbol \Psi} }}
\newcommand{\epsilonpol}{\ensuremath{ {\boldsymbol \epsilon}^{(\pol)}  }}
\newcommand{\spin}[1]{\mbox{spin~$#1$}}

\newcommand{\tin}{\ensuremath{t=0 } }

\title[Scaling relations \& dynamical heating in ULDM]{Scaling relations, dynamical heating and tidal disruption in spin~$s$ ultralight dark matter models}

\author[López-Sánchez et al.]{
Jessica N. López-Sánchez,$^{1}$\thanks{E-mail: lopez@fzu.cz}
Erick Munive-Villa,$^{1}$\thanks{E-mail: munive@fzu.cz}
Constantinos Skordis$^{1,2}$
Federico R. Urban$^{1}$\thanks{E-mail: federico.urban@fzu.cz}
\\
$^{1}$CEICO—FZU, Institute of Physics of the Czech Academy of Sciences, Na Slovance 1999/2, 182 00 Prague, Czech Republic\\
$^{2}$Department of Physics, University of Oxford, Denys Wilkinson Building, Keble Road, Oxford OX1 3RH, UK \\
}

\date{Accepted 2025 September 18. Received 2025 September 18; in original form 2025 February 17}

\pubyear{2025}

\begin{document}
\label{firstpage}
\pagerange{\pageref{firstpage}--\pageref{lastpage}}
\maketitle

\begin{abstract}
We explore the impact of \spin{0}, \spin{1} and \spin{2} ultralight dark Matter (ULDM) on small scales by numerically solving the 
Schr\"odinger-Poisson system using the time-split method. We perform simulations of ULDM for each spin, starting with different 
numbers of identical initial solitons and analyse the properties of the resulting haloes after they merge. Our findings reveal that 
higher spin lead to broader, less dense haloes with more prominent Navarro-Frenk-White (NFW) tails, a characteristic that 
persists regardless of the number of solitons involved. Additionally, we study the process of dynamical heating for these haloes, 
and find that the heating time-scale for higher spin increases order an of magnitude compared to the \spin{0} case. Then, we identify 
scaling relations that describe the density profile, core-NFW of spin~$s$ ULDM haloes as a function of the number of initial 
solitons $\Nsol$. These relations allow us to construct equivalent haloes based on average density or total mass, for arbitrarily large 
$\Nsol$, without having to simulate those systems. We simulate the orbit of an ULDM satellite in a constructed halo treated as an 
external potential, and find that for host haloes having the same average density, the disruption time of the satellite is as predicted 
for a uniform sphere regardless of the spin. However, satellites orbiting haloes having the same mass for each spin, result in faster 
disruption in the case of \spin{0}, whereas for haloes having the same core size result in faster disruption in the case of \spin{2}.
\end{abstract}




\section{Introduction}

Ultralight dark matter (ULDM), namely bosonic dark matter particles whose mass is of order \(10^{-22}\,\text{eV}\), has been established as a viable and phenomenologically rich candidate for the observed cosmological dark matter~\citep{Niemeyer-review}. ULDM is modelled as an oscillating classical field minimally coupled to gravity, existing as a superposition of nearly coherent waves, with \spin{0} (scalar field)~\citep{Ferreira_2021, Hu-2000,Matos:1999et}, \spin{1} (vector field) or \spin{2} (tensor field)~\citep{Jain_2022,Alexander_2021}. In these models, provided the mass is sufficiently small, the de Broglie wavelength is of the order of kiloparsecs, the typical size of observable galaxies in the Universe. The result is an effective 'quantum pressure' that counteracts gravitational attraction which then has an impact on the formation and distribution of structures at small scales.

ULDM may be compared to cold dark matter (CDM) which is a collision-less cold fluid that forms self-bound, virialised units called haloes through a hierarchical process. Both ULDM and CDM predict the formation of large-scale structures in the Universe in concordance with observations from cosmological surveys at large scales. However, ULDM may have an edge when confronted with observations at small scales, where CDM predictions seem to be in tension with the data~\citep{smallscaleprob1,smallscaleprob2}---see also~\citep{Feng_2010, Bertone_2018,TULIN20181} for an overview of the alternatives.

While \spin{0} ULDM has been the subject of investigation over the last two decades, the study of higher spin ULDM using both analytic and numerical methods is more recent. 
Small-scale simulations of solitonic configurations for \spin{0} and \spin{1} ULDM were contrasted in~\citet{Amin_2022}, where it was shown that the central region of solitons in \spin{1} ULDM is less dense and has a smoother transition as the radius increases compared to the \spin{0} case. Additionally, it was found that solitons for \spin{1} and \spin{2} are formed later than for the \spin{0} case, that is, the higher the spin, the larger the soliton condensation time \citep{Jain:2023ojg}. In all cases, the solitons are surrounded by a Navarro-Frenk-White (NFW) envelope connected to other filamentary structures~\citep{Gorghetto_2022,Jain:2023ojg,chen2023gravitational}. These results show differences between each model in simple configuration ensembles, which can give rise to significant observable effects. Two such effects concern the dynamics of satellite subhalo systems within a host halo, specifically, their tidal disruption and the effect on dynamical heating.

The tidal disruption of subhaloes has been extensively explored within the CDM model. In \citet{hui2017ultralight} it is shown that, contrary to what would happen with an ULDM structure, the subhaloes in CDM would orbit forever, provided that they are within the tidal radius in the absence of dynamical friction. Furthermore, \citet{van2018disruption} reported that physical disruption of CDM subhaloes with NFW profiles is a relatively rare occurrence. However, these subhaloes do experience mass loss, which depends on the amount of energy injected into them. In fact, the density profiles of self-bound remnants are fully determined by the fraction of mass lost, and tend to approach an exponentially truncated NFW profile \citep{errani2021asymptotic}.

In contrast to CDM, disruption always occurs in the case of ULDM, as this type of dark matter can eventually tunnel through the potential barrier given a sufficiently long timescale \citep{hui2017ultralight}. 
This phenomenon has been studied only in the \spin{0} case. In~\citet{hui2017ultralight}, the tidal radius of a \spin{0} ULDM satellite was estimated using a spherically symmetric tidal potential $\propto r^2$ (centred around the satellite) using the time-independent Schr\"odinger-Poisson system (SP). It was shown that \spin{0} ULDM within the tidal radius can escape to infinity by tunnelling through the potential barrier at the tidal radius, implying that all systems subjected to an external tidal field will eventually be disrupted. The survival time of a satellite subhalo depends on the ratio of its central density to the average density of host halo over the orbital radius of the satellite. It was found that larger such ratios result in more circular orbits before disruption happens. The time-independent approximation to the SP system was questioned in~\citet{du2018tidal} where it was seen to be valid only for small enough times. Considering time dependence and still within the $\propto r^2$ tidal potential model, it was found that the core loses mass faster and becomes increasingly susceptible to tidal effects, leading to faster disruption times. Increasing the model complexity,~\citet{du2018tidal} also performed full three-dimensional numerical simulations for determining the time-dependent profile of a \spin{0} ULDM satellite moving in a host halo modelled as a uniform sphere with a fixed mass and treated as an external potential. In this case, the satellite loses mass gradually and quickly relaxes to a less compact configuration, which can be described by a new soliton with lower central density. Using their numerical simulations,~\citet{du2018tidal} then estimated the survival time of satellite galaxies in the Milky Way.

The effects of tidal disruption in the context of spin-0 ULDM was further studied in \citet{TidalSelf} with the inclusion of both attractive and repulsive self-interactions. It was found that repulsive interactions enhance the efficiency of disruption, whereas attractive interactions have the opposite effect. These phenomena are particularly relevant because it is possible to reproduce the effects of a self-interacting scenario by adding extra degrees of freedom to the model, as in the case of spin-$s$ ULDM, where solitons would become more or less likely to disrupt depending on the spin value \citep{Amin_2022,Gosenca:2023yjc,Jain:2023ojg}

The study of the dynamics between satellite galaxies and their host haloes within alternative dark matter models and examining their effects on survival time, structural configurations and mass transfer, can reveal significant differences that can be compared with observational data. Specifically, one may ask how the satellite dynamics changes for ULDM models with different spin $s$. In this work, we investigate the properties of haloes formed through the mergers of soliton configurations, characterising their density profiles using universal scaling relations. We then apply these findings to model the dynamics of a satellite within a realistic external potential that we numerically compute for a spin~$s$ ULDM halo using the SP system. By considering \spin{0}, \spin{1} and \spin{2} ULDM, we aim to distinguish the effects specifically attributed to having additional degrees of freedom arising from the different ULDM spins.

The paper is organised as follows: in Section \ref{sec: sol_basis}, we discuss the non-relativistic modelling of ULDM, focussing on the multi-component Schr\"odinger-Poisson system for \spin{0}, \spin{1} and \spin{2}. Section \ref{sec: code} provides a brief overview of the numerical methods employed to evolve the system of equations in each model. In Section \ref{sec: MBMergers} we report on the results of multiple soliton mergers of each spin $s$ model in order to explore the evolution and properties of the resultant halo, including the density profile, total energy and spin density. In section~\ref{sec: URSDP}, we identify scaling relations between the initial density profile of the solitons and the density profile of the final soliton, corresponding to a final halo profile, created through their merging. This allows us to construct ULDM haloes with the equivalent properties for each model. In section~\ref{sec: Sathaloes}, we apply the resultant dark matter profiles to use them as host haloes of satellite systems. In this case, the host halo is made of \spin{0}, \spin{1} or \spin{2} and is considered an external potential. We summarise our conclusions in Section \ref{sec: Conclu}.

\section{Non-relativistic approach for spin~$s$ ULDM}\label{sec: sol_basis}

A spin~$s$ massive field in the non-relativistic limit can be described by the multiple-component SP system~\citep{Jain_2022, Adshead_2021}
\begin{align}\label{eq:multipleSP}
    \begin{aligned}
    i\hbar\frac{\partial}{\partial t}\Psivec&=-\frac{\hbar^2}{2m_s }\nabla^2\Psivec+m_s\Phi\Psivec,\\
    \nabla^2\Phi&=4\pi G\rho_{0} ( \text{Tr}[\Psivec^\dagger\Psivec]-1),
    \end{aligned}
\end{align}
where $\hbar$ stands for the reduced Planck constant, $G$ is the gravitational constant, $m_s$ is the mass of the ULDM particle and $\rho_0$ is the mean density of the simulation. We have normalised the wavefunction to the mean density on the simulation so that $\langle\text{Tr}[\Psivec^\dagger \Psivec]\rangle = 1$ and $\text{Tr}[\Psivec^\dagger\Psivec]$ is the probability density of occupation. Throughout this section, we use the notation in~\citet{Jain_2022} to construct the initial conditions of the simulations. Then, the field is expressed as a function of the spin as follows:
 \begin{align}\label{eq: components}
    \begin{aligned}
    \psi&=[\Psivec] \quad &\text{spin~0},\\
    \psi_i&=[\Psivec]_i \quad &\text{spin~1},\\
    \psi_{ij}&=[\Psivec]_{ij} \quad &\text{spin~2}.
    \end{aligned}
\end{align}

The trace is defined as $\text{Tr}[\Psivec\Psivec^\dagger] = \psi_i\psi_i^\dagger$ and $ \text{Tr}[\Psivec\Psivec^\dagger]= \psi_{ij}\psi_{ji}^\dagger$ for \spin{1} and \spin{2}, respectively. A massive spin~\textit{s} field admits $2s+1$ spin configurations characterised by the orthogonal set $\{\epsilonpol\}$, where $\pol \in \{-s,\ldots ,s\}$ is the polarisation. Then, the spin~$s$ wave function can be decomposed as
\begin{equation}\label{eq:multiplicitystates}
    \Psivec(t,\xcoord)=\sum_{\pol}\psi_\pol(t,\xcoord)\epsilonpol,
\end{equation}
where $\psi_\pol$ is the field with polarisation $\pol$. 

In what follows we are interested in setting up the system as being composed of multiple spin $s$ solitons in the ground state, and letting them evolve in time according to \eqref{eq:multipleSP}. For the \spin{0} case, the ground state $\psi_\text{sol}$ is a real function that satisfies the time-independent SP system~\citep{Guzman-2004}, as described in Appendix~\ref{a: GS-soliton}. Without loss of generality, for higher spins we may take the ground-state of each soliton at the initial time $\tin$ to be as in the \spin{0} case equal to $\psi_\text{sol}$, multiplied by a real coefficient $c_{\pol}$ and a phase $\theta_\pol$, such that
\begin{equation}\label{eq:polarisedwf}
    \psi_\pol\left(\tin,\xcoord\right)= \psi_\text{sol}\left(\xcoord\right) c_\pol \text{e}^{-i\theta_\pol}.
\end{equation}
We assign the coefficients $c_{\pol}$, which determine the mixing amongst the $2s+1$ spin configurations and satisfy $\sum_{\pol}c_{\pol}^2=1$, and  the phase  $\theta_\pol \in [0,2\pi)$, randomly for each soliton. A detailed description of how to compute the spherical ground state solution $\psi_\text{sol}$ can be found in Appendix~\ref{a: GS-soliton}. 

\subsubsection*{Spin~0 }

This case is the simplest, and the field is defined by $\psi_{\pol}$ in equation \eqref{eq:polarisedwf} with \(c_p=1\):
\begin{equation}\label{spin0}
    \psi_0 =  \psi_\text{sol} \text{e}^{-i\theta_0}.
\end{equation}

\subsubsection*{Spin~1 }\label{sec: spin1}
The basis is represented by the following set of orthonormal vectors, associated with three polarisation states $\pm 1$ and 0~\citep{Jain_2022}:
\begin{equation}\label{eq:Base1}
    \boldsymbol\epsilon^{(\pm 1)} \equiv 
   \frac{1}{\sqrt{2}}\begin{pmatrix}
        1\\\pm i\\0
    \end{pmatrix}; 
\quad \boldsymbol\epsilon^{(0)}\equiv 
   \begin{pmatrix}
        0\\0\\1
    \end{pmatrix}.
\end{equation}
We assign the two $c_{\pol}$ coefficients randomly and determine the third using $\sum_{\pol}c_{\pol}^2=1$ for each constructed soliton according to~\eqref{eq:polarisedwf}. This is equivalent to constructing orthogonal random vectors.

\subsubsection*{Spin~2 }\label{sec: spin2}
In this case, five independent states are defined by the polarisation $\pm 2, \pm1$ and 0. The maximally polarised orthonormal tensors can be written in terms of the following orthogonal and traceless matrices 
\begin{align}\label{eq:Basis2}
\begin{aligned}
        \boldsymbol{\epsilon}^{(\pm 2)}&\equiv \frac{1}{2}\begin{pmatrix}
        1 & \pm i & 0\\
        \pm i & -1 & 0 \\
        0 & 0 & 0
    \end{pmatrix},\\
    \boldsymbol{\epsilon}^{(\pm 1)}&\equiv \frac{1}{2}\begin{pmatrix}
        0 & 0 & 1\\
        0 & 0 & \pm i \\
        1 & \pm i & 0
    \end{pmatrix},\\
     \boldsymbol{\epsilon}^{(0)}&\equiv\frac{1}{\sqrt{6}}\begin{pmatrix}
        -1 & 0 & 0\\
        0 & -1 & 0 \\
        0 & 0 & 2
    \end{pmatrix}.
\end{aligned}
\end{align}
This case has five  $c_{\pol}$ elements which are again assigned randomly subject to $\sum_{\pol}c_{\pol}^2=1$  for each soliton.

Let us note that since we have ignored the possible self-interactions, whose structure is different for each type of spin, the dynamical equations \eqref{eq:multipleSP} 
can be  mathematically mapped to a multi-field system of $n_s$ scalars. For the spin cases that we have studied, this corresponds to $n_s=1$, $n_s=3$ and $n_s=5$.
Such multi-field systems have been studied in~\citet{Gosenca:2023yjc}, however, in the $n_s=1$, $n_s=2$ and $n_s=4$ cases. While the evolution equations are equivalent,
it is expected that the self-interactions will induce important differences in the early universe when these fields are produced, and this will manifest into the initial conditions.
Specifically, in the multi-field case, one would expect that each initial soliton will have rougly similar distribution of each constituent field with some fluctuations, resulting in almost zero spin density
and in~\citet{Gosenca:2023yjc} it was chosen to be the same for all field components. 
The probability of extreme cases where different solitons have very different field-composition is statistically suppressed and not expected to occur. 
On the contrary, such cases are possible in the spin~$s$ cases, where any field configuration amongst the spin basis can occur; see~\citet{Jain_2022}
 Indeed, specific mechanisms that change the relative amplitudes of the field components over time have been proposed, for example in~\citet{amaral2024vectorwavedarkmatter,PhysRevD.111.103520}
 In our case we have chosen a random polarisation drawn isotropically from the 2$s$-sphere.
 This difference leads to important physical effects that manifest during mergers, resulting in distinct post-merger density profiles compared to the multifield case.
In Appendix~\ref{sec: spinVsMulti} we present a comparison between the resultant halo densities for the \spin{1}  and the $n_s=3$ multi-field  cases.

We note also that one could in principle choose a different basis than the spin representation \eqref{eq:Base1} and \eqref{eq:Basis2}, however,
the decomposition in terms of the given spin basis makes the connection with the underlying physics models clearer, see~\citet{Jain_2022}
It is trivial to transform  the \spin{1} basis \eqref{eq:Base1}  to a (multi-field) basis of three unit column vectors via a linear transformation, 
however, transforming the  \spin{2} case into $5$ column unit vectors, rather than \eqref{eq:Basis2} is less so and makes the underlying physics obscur.
 Moreover, with the introduction of self-interactions,  it is expected that the spin bases will be better suited, particularly in the case of \spin{2}.

\section{Numerical implementation}\label{sec: code}

We have developed a new numerical code in C++ which solves the SP system~\eqref{eq:multipleSP} using a time-splitting pseudospectral method. For the systems under study, Fourier methods perform better than numerical local methods since the complexity is of the order of $N \text{log}_2 N$, whereas the Finite Difference Method or the Finite Element Method has a complexity of the order of $N^2$, being $N$ the total number of operations required for each time-step~\citep{numrecip}. This numerical technique has also been implemented in other works to study the evolution of the scalar field, such as~\citet{Edwards_2018,volkerLarge}. 

In this method, the time step $\Delta t$ is expressed as a combination of operations in configuration and in Fourier space, which are applied to each component of the spin~$s$ system \eqref{eq: components}, considering that there are three and five independent terms for \spin{1} and \spin{2}, respectively. Specifically, starting from the wavefunction $\psi_\pol\left(t, \mathbf{x}\right) $ for each component $\pol$ at time $t$,  we first compute the $\psi_\pol\left(t + \Delta t/2, \mathbf{x}\right) $ for each $\pol$ at the half time-step $\Delta t/2$. All wavefunction components are then used 
for evaluating the potential $\Phi(t + \Delta t)$ by solving the Poisson equation. We finally combine both steps to evaluate the wavefunction $\psi_\pol\left(t + \Delta t, \mathbf{x}\right)$, as is captured by the following set of equations
\begin{subequations}
\label{eq: kdk}
\begin{align}
    \psi_\pol\left(t+\Delta t/2\right) =& \mathcal{F}^{-1}\left[
    \text{e}^{- \frac{ i \Delta t \tilde\hbar \kvec^2}{2 m_s }}
    \mathcal{F}  \left(\text{e}^{-\frac{i \Delta t \Phi(t)}{2 \tilde\hbar}} \psi_\pol(t)\right)\right],
\label{psi_pol_delta_t_2}
\\
    \Phi\left(t+\Delta t\right) =& \mathcal{F}^{-1}\left[-\frac{1 }{\kvec^2} \mathcal{F} \left( 4\pi G \rho_0 ( \text{Tr}[\Psivec^\dagger\Psivec]-1)\right) \right],
\label{Phi_delta_t}
\\
    \psi_\pol\left(t + \Delta t\right) =& \text{e}^{ -\frac{i\Delta t \Phi(t+\Delta t)}{2 \tilde\hbar}}   \psi_\pol\left(t+\Delta t/2\right) ,
\label{psi_pol_delta_t}
\end{align}
\end{subequations}
where $\kvec$ is the spatial frequency domain, $\mathcal{F}$ and $\mathcal{F}^{-1}$ respectively stand for the Discrete Fourier Transformation and its inverse, and $\text{Tr}[\Psivec^\dagger\Psivec] = \sum_\pol \psi_\pol \psi_\pol^\dagger$, evaluated at $t + \Delta t/2$. This works because from  \eqref{psi_pol_delta_t} we have that $|\psi_\pol(t+\Delta t)|^2 = |\psi_\pol(t+\Delta t/2)|^2$ so that  \eqref{Phi_delta_t} can be consistently used. The error associated with this numerical approach is of order $\mathcal{O} (\Delta t^3)$~\citep{PyUltraLightSI}.

To be able to simulate maximum ULDM velocities of $v_\text{max} \sim 100$~km/s for a given N-body simulation box size and ULDM mass, a minimum number of grid points is required. 
For example, for a mass value of $m_s = 2.5\times10^{-22}$~eV, at least $415^3$, $4150^3$ and $41500^3$ points are needed for volumes of $0.1$, $1$ and $10$~Mpc$^3$, respectively. For lighter masses such as $m_s = 1.75\times10^{-23}$ eV, these values decrease significantly, reaching a minimum requirement of $2900^3$ points even in a modest simulation size of 10~Mpc$^3$. Thus, the computational power required to run large-scale simulations becomes evident.

To obtain a robust numerical solution it is necessary to resolve the structures on the scale of the de Broglie wavelength $\lambda_\text{dB}=\displaystyle\frac{\tilde\hbar}{v}$ where $\tilde\hbar=\displaystyle\frac{\hbar}{m_s}$ and $v$ is an estimate of the velocity of an ULDM fluid packet. We can obtain $v$ by appealing to the Madelung representation which gives $v = \tilde\hbar \left|\nabla\alpha\right|$, where $\alpha$ is the phase of the wave function $\psi$ and ranges between $[0,2\pi]$. Assuming a half-step approximation for the spatial derivative, the maximum velocity that this method can resolve is estimated as $v_\text{max} \sim \displaystyle\frac{\pi\tilde\hbar}{\Delta x}$~\citep{volkerLarge}, leading to the resolution criterion
\begin{equation}\label{eq:minRes}
    \Delta x < \frac{\pi \tilde\hbar}{v_\text{max}}.
\end{equation}
Then, following the Courant-Friedrich-Lewy condition~\citet{Courant1928} for parabolic equations and considering that the phase of the wave function expressed in~\eqref{eq: kdk} should be smaller than $2\pi$, $\Delta t$ must fulfil the condition
\begin{equation}\label{eq:dt}
    \Delta t < \min \left( \frac{4 \Delta x^2}{3\pi\tilde\hbar} ,  \frac{2\pi \tilde \hbar}{|\Phi|_{\max}} \right).
\end{equation}

All our simulations were performed in a cubic box of $L = 100$kpc. To ensure good convergence we used a mesh of $\Ngrid=512^3$ grid points in the case of \spin{1} and \spin{2} with $\Delta x = 0.195$ kpc, and $v_\text{max} \sim 123$km/s, as given by \eqref{eq:minRes}. However, as we discuss in appendices~\ref{sec: stability} and~\ref{sec: lambda_conv}, we found that when merging a large number of initial solitons this is not sufficient in the  \spin{0} case, but using $\Ngrid=1024^3$ convergence was indeed reached (corresponding to $\Delta x = 0.098$kpc and $v_\text{max} \sim 247$km/s). This is because the cores are more compact in the \spin{0} case and thus closer to the resolution limit compared to \spin{1} and \spin{2} cases where the density is distributed across two and five independent components, respectively. An analysis of the stability criteria of these configurations using different resolutions is provided in Appendix~\ref{sec: stability}.
Finally, we accelerated our simulations using the Fast Fourier Transformation library implemented in CUDA (cuFFT\footnote{\url{https://docs.nvidia.com/cuda/cufft/index.html}}) for general computing on graphical processing units (GPUs). In Appendix~\ref{sec:Performance}, we discuss briefly the performance enhancement when using GPUs.

\section{Multiple soliton merger}\label{sec: MBMergers}

We consider an idealised initial configuration of solitons for each spin~$s$ case. These solitons have identical masses and radial profiles but different spin distribution chosen randomly and drawn  isotropically from the 2$s$-sphere as 
described in Sec.\ref{sec: sol_basis}.

The number of initial solitons, $\Nsol$, is varied as a proxy for the halo mass, which allows us to focus specifically on the impact of spin on the halo formation process, 
following a similar methodology to that described in~\citet{Jain_2022}.
Such idealised setups are an important step in developing fully fledged spin-$s$ simulations which also include the self-interactions,
as they can provide a basis of comparison between the underlying spin model.
A realistic cosmological scenario would of course involve not only solitons of different masses but also isolated solitons merging with haloes at different stages in their evolutions, as well as, other processes such as accretion of material
and baryons.  
Nonetheless, our controlled setup  allows us to directly track macroscopic differences that arise solely because of the ULDM spin.

\subsection{Initial conditions}
\label{sec: ics}
We ran 24 simulations for each spin~$s$ ULDM model, varying the number of solitons, $N_\text{sol}$, in steps of $5$, starting from $5$ up to $120$. The solitons were initially positioned randomly within the subdomain $[10, 90]$ kpc, so they were sufficiently far from the boundaries, and with zero linear momentum. Each soliton configuration was generated numerically following the procedure outlined in Appendix \ref{a: GS-soliton}, with a fixed scaling factor $\lambda = 1000$ and a scalar field mass $m_s = 2.5 \times 10^{-22}$eV. In the  \spin{0} case, each soliton is described as in~\eqref{spin0} with the phase assigned randomly.  For \spin{1} and \spin{2}, each soliton is partially polarised through the linear combination given by~\eqref{eq:multiplicitystates} in terms of the set of maximally polarised basis defined by \eqref{eq:Base1} and \eqref{eq:Basis2}, respectively. The coefficients $c_{\pol}$ have been assigned randomly for each soliton following \eqref{eq:polarisedwf}. The mass of the soliton was computed in terms of the integrated density out to infinity, $M=\int_V\rho dV$, in isolation. Thus, all solitons have the same  mass $M_\text{sol} = 5.31 \times 10^7 M_{\odot}$, but different polarisations $c_{\pol}$ and phases $\theta_\pol$ in their wavefunction. 

\begin{figure}
    \centering
    \includegraphics[width=0.9\linewidth]{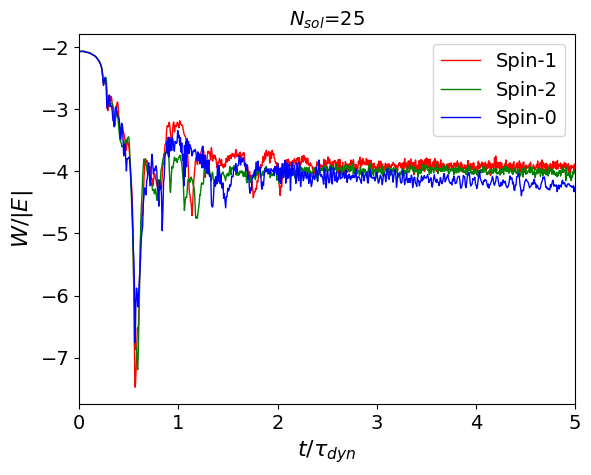}
    \caption{Evolution of the ratio $W/\abs{E}$ as a function of $t/\tau_{\text{dyn}}$ for \spin{0}, \spin{1} and \spin{2} models with $N_\text{sol}=25$.}
    \label{fig: virial_energy}
\end{figure}
\begin{figure*}
    \centering
    \includegraphics[width=0.75\textwidth]{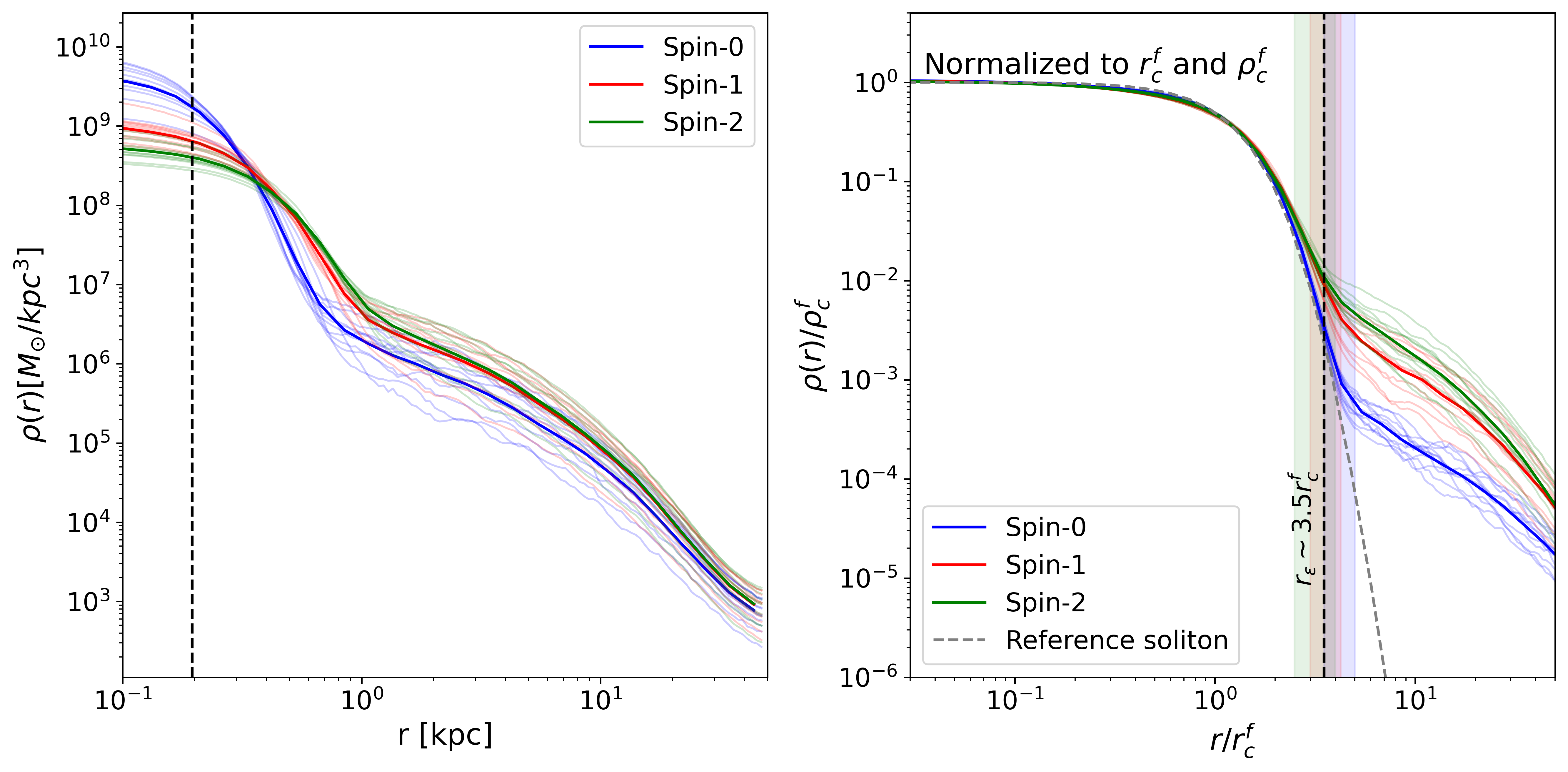}    
    \caption{\textbf{Left panel}. Density profiles for each type of \spin{s} simulation in the range $10<N_\text{sol}\leq120$, constructed via spherical averaging. The thin curves denote the density profile for a given $N_\text{sol}$ and different spin-$s$. The solid lines correspond to the radial average of each model and are sorted from top to bottom at small radii for \spin{0} (blue), \spin{1} (red), and \spin{2} (green). The vertical black dashed line represents the spatial resolution. The mass of the ULDM particles is $m_s=2.5\times 10^{-22}$ eV, and the mass of each soliton is $M=5.31\times 10^7 M_{\odot}$. \textbf{Right panel}. Density profile normalised by the maximum density value $\rho_c^f$ as a function of the radius normalised by $r_c^f$. The reference soliton configuration (using $r_c = \rho_c =1 $ in~\eqref{eq: join_profile}) is shown for comparison. The ordering at large radii is inverted.}
    \label{fig:DMPRofMM}
\end{figure*}
\begin{figure}
    \centering
    \includegraphics[width=0.85\linewidth]{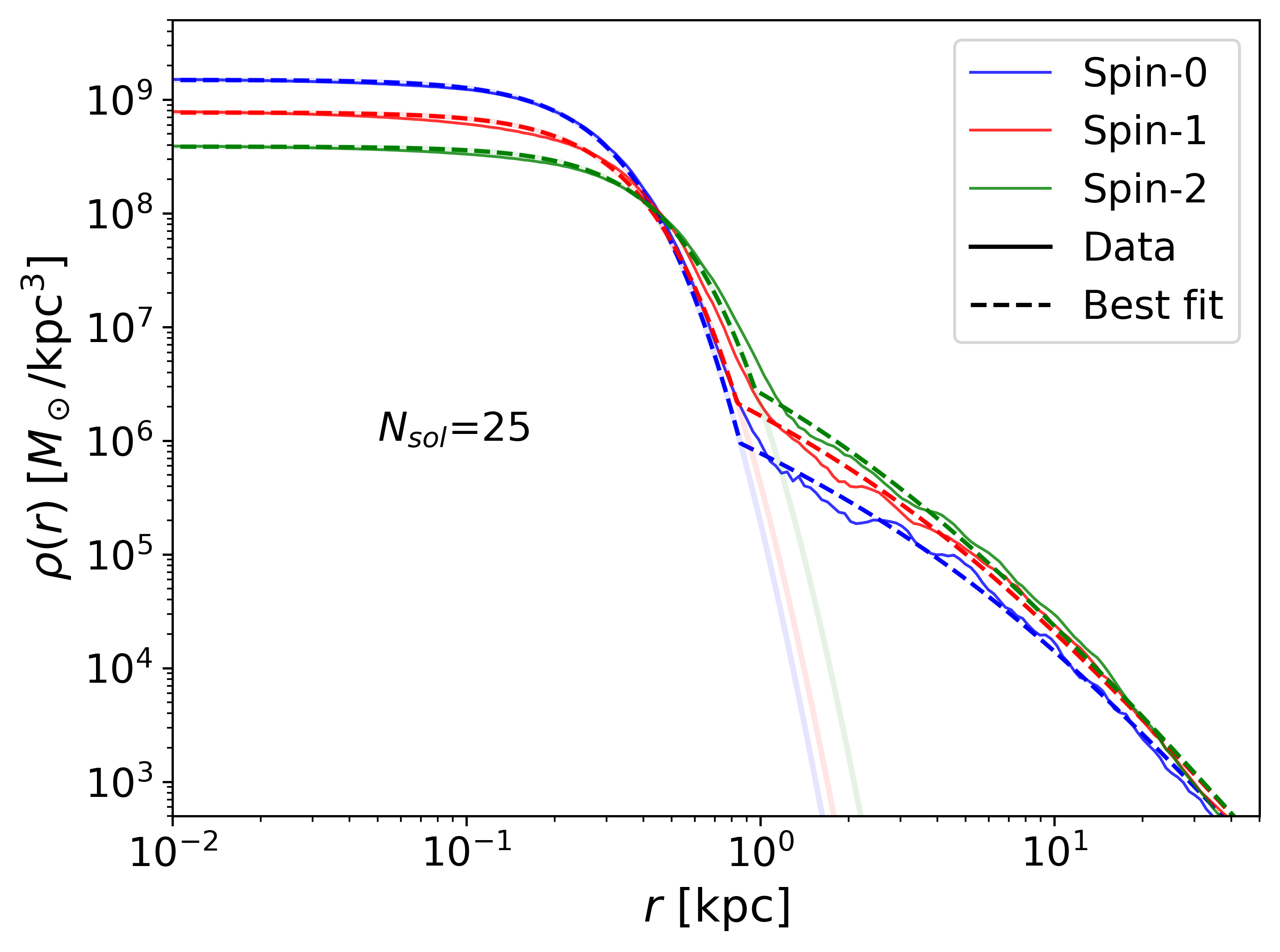}
    \caption{The solid lines represent the density of dark matter computed directly from simulations, showing the $\Nsol = 25$ case. The dashed lines show the best fits obtained from Eq.~\eqref{eq: sol} and Eq.~\eqref{eq: nfw} and the scaling relations discussed in Sec.~\ref{sec: URSDP}. The solitonic cores are displayed with fainter lines for comparison.}
    \label{fig:FitsProf}
\end{figure}
\subsection{Evolution} 
\subsubsection{Choosing the stopping time}

For studying the final steady-state configurations, the system was allowed to evolve until a final time of  $t_f = 20\tau_{\text{dyn}}$, where the dynamical time $\tau_{\text{dyn}}$ is defined as
\begin{equation}
    \tau_{\text{dyn}} = \displaystyle\frac{1}{\sqrt{G\rho_0}}.
\end{equation}
and denotes the typical timescale that a system needs to relax to an equilibrium configuration when collapsing under gravity. Here, $\rho_0$ is the mean density from eq. \eqref{eq:multipleSP}.
We show in section~\ref{sec: energies}  that the system virialises after $t\sim 2\tau_{\text{dyn}}$ for each model and so taking  the final configuration at $t_f= 20\tau_{\text{dyn}}$ is justified as it is more probable  to lead to a stable and virialised system.

If the initial number of solitons is smaller than $N_\text{sol}<5$ then $\tau_\text{dyn}>30$ Gyr which makes the simulations computationally demanding, as we verified explicitly. We thus focus our analysis to $N_\text{sol}\geq10$. Moreover, if $N_\text{sol}>120$, then the spatial resolution and box size we use would not be enough, which sets our choice of $N_\text{sol}\leq120$.

\subsubsection{Energy evolution}\label{sec: energies}
The stability criterion of the final halo configuration in each model can be studied using the quotient $W/\abs{E}$, where $E = W+K$ is the total energy of the system, $K$ the kinetic energy and $W$ the potential energy, defined as~\citep{Jain:2023ojg}
\begin{align}\label{eq:Energy}
    K &=\frac{\tilde\hbar^2}{2}\int_V dV \text{Tr}[\nabla\Psivec^{\dagger}\cdot \nabla\Psivec], 
\\\label{eq:Energy2}
    W &= \frac{1}{2}\int_V dV \Phi\text{Tr}[\Psivec^{\dagger}\Psivec].
\end{align}

Fig.~\ref{fig: virial_energy} shows the evolution of $W/\abs{E}$ as a function of $t/\tau_{\text{dyn}}$ for each spin with $\Nsol =25$, chosen without loss of generality.\footnote{In Appendix \ref{sec: stability}, Fig.~\ref{fig:MergeTime}, we show the evolution of the potential energy as a function of the dynamical time for different values for $N_{\text{sol}}$. The merger process occurs around $0.6\tau_{\text{dyn}}$ in all cases; after that the system gradually relaxes.} 
Note the initially large energy fluctuations, particularly around $t/\tau_{\text{dyn}}\sim 1\, -\, 2$. The relaxation process can, however, last for hundreds of $\tau_\text{dyn}$.
 We observe that before the merger, the potential energy dominates, reaching a maximum value when the collision starts. After that, the three spin~$s$ models converge to a roughly constant value, with \spin{0} having a slight slope. This plot demonstrates that the system slowly stabilises to a specific value of $W/\abs{E}$, which then remains approximately constant over time. We checked that the remaining simulations with different $\Nsol$ exhibit similar behaviour. We note that the asymptotic value of $W/\abs{E}$ depends also on the spin content of the initial solitons and so nothing can be said about the hierarchy observed in Fig.~\ref{fig: virial_energy} between the three spins. We have checked that other $\Nsol$ cases show a  different hierarchy. 

\subsection{Properties of the resulting profiles}

\subsubsection{Resulting density profile}\label{sec: dens}
If the halo resulting from the soliton mergers is approximately spherically symmetric, we can average its density over concentric spheres. These spherical averages are shown on the left panel of Fig.~\ref{fig:DMPRofMM}. We then fit these averages into template functions adopting the prescription of \spin{0} ULDM dark matter haloes as in~\citet{schive2014cosmic,Schive:2014hza},
\begin{equation}
    \rho_{\text{halo}}(r)=\Theta(r_{\epsilon}-r)\rho_{\text{sol}}(r)+\Theta(r-r_{\epsilon})\rho_{\text{NFW}}(r),
\label{eq: join_profile}
\end{equation}
to fit the halo density profile in each spin~$s$ model and final soliton configuration. Here, $\Theta$ is the step function, and $r_{\epsilon}$ is the transition radius between the soliton and the NFW tail, which are described by the following expressions
\begin{align}
\label{eq: sol}
    \rho_{\text{sol}}(r)=&\frac{\rho_c}{\left[1+\alpha(r/r_c)^2\right]^8},
  \\ 
  \label{eq: nfw}
    \rho_{\text{NFW}}(r)=&\frac{\rho_s}{\left(r/r_s\right)\left(1+r/r_s\right)^2},
\end{align}
respectively, where $\alpha = 0.091$ was fixed as in \citet{Schive:2014hza}. We defined the centre of the final solitonic core as the point with the maximum density, and used this density as the parameter $\rho_c^f$, where the superscript $f$ denotes ``final soliton''. The parameter $\rho_s$ is determined through $\rho_{\text{sol}}(r_\epsilon) = \rho_{\text{NFW}}(r_\epsilon)$, so we are left with three parameters to fit: $r_\epsilon$, $ r_c^f$ and $r_s$. However, naively doing this does not take into account the abrupt change from a core to an NFW profile which occurs at $r_{\epsilon}$. This creates a strong degeneracy between $r_c^f$ and $r_\epsilon$ leading to best fits which under or over-predict the profile for a range of radii in the immediate neighbourhood of $r_\epsilon$ and which are visibly distinguishable from the averaged profile. Thus our strategy was to first fit the final solitonic core (by cutting off the NFW tail) to \eqref{eq: sol} with a single parameter $r_c^f$, and only then fit the total profile by varying only $r_\epsilon$  and $r_s$. For our fits, we run Monte Carlo Markov Chain chains using the Affine Invariant Ensemble Sampler~\citep{Foreman-Mackey:2012any}. The fits for the case of $N_\text{sol}=25$ are shown in Fig.~\ref{fig:FitsProf}. 
\begin{figure}
    \centering
    \includegraphics[width=0.38\textwidth]{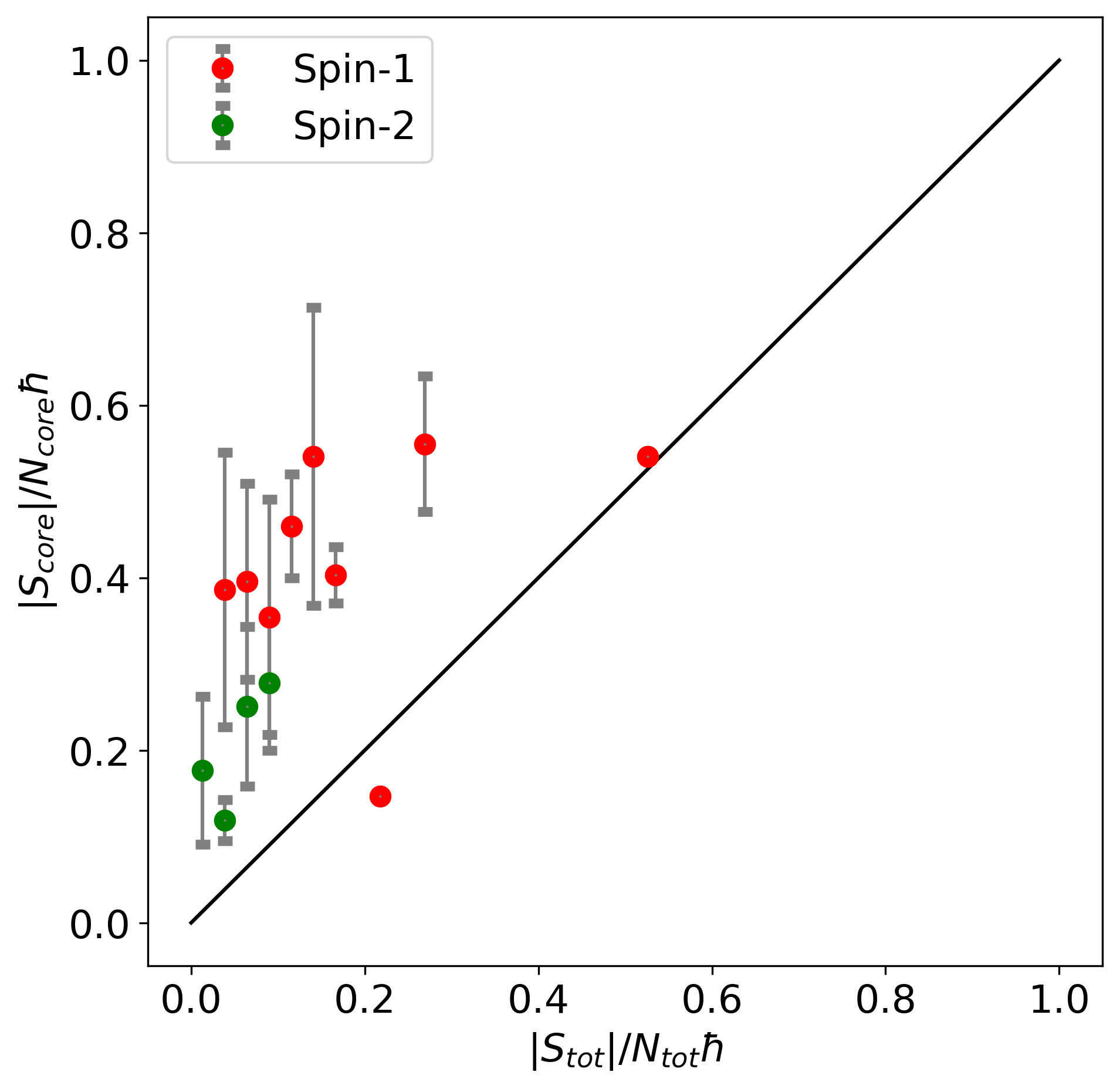}  
    \caption{ 
The normalised core spin $\boldsymbol S_{\text{core}}/\Ncore$ versus normalised total spin $\boldsymbol S_{\text{tot}}/\Ntot$. The bars indicate the standard deviation for all the simulations, and the points represent the binned average data. The trianles correspond to \spin{1} and solid circles to \spin{2}.}
    \label{fig:Spin}
\end{figure}
In Fig.~\ref{fig:DMPRofMM} (left panel) we display the set of final halo density profiles for \spin{0} (in blue), \spin{1} (in red) and \spin{2} (in green). The thin lines depict the density profile from simulations with different $N_\text{sol}$, the thick lines mark  the average profile for each spin, and the vertical black dashed line denotes the numerical resolution. We observe that, within our chosen range of solitons $10<N_\text{sol}\leq120$ the profiles corresponding to each model exhibit similar behaviour, which can be effectively described by the average profile, a fact which has potential use in observational comparisons. A noticeable difference in the shape of the profiles is observed in the central regions between \spin{0} and \spin{1}, consistent with the findings of ~\citet{Amin_2022}. In contrast, the difference in central density between \spin{1} and \spin{2} is less pronounced. Increasing the spin leads to less pronounced interference patterns (see section~2.2 of~\citet{Amin_2022}) since the probability of constructive interference decreases with higher spin.\footnote{This is akin to what is observed in the \spin{0} multi-field case, see~\citet{Gosenca:2023yjc}} As the radius increases, \spin{1} and \spin{2} exhibit a smoother transition than the \spin{0} case and the density profiles for all spins  converge together at larger radii, as expected.

In Fig.~\ref{fig:DMPRofMM} (right panel) we show the final density normalised with the maximum density value $\rho_c^f$, as a function of the radius normalised with $r_c^f$. Once more, the thin lines represent the scaled densities from the simulations within the set $10<N_\text{sol}\leq120$, while the dark lines mark the average density for each spin. The differences in the density profile tails are now more pronounced, with  the  transition from the soliton to the NFW tail being sharper for \spin{0} and becoming increasingly smoother with increasing spin. We observe that the transition radius $r_{\epsilon}$ for these final haloes lies in the range $r_{\epsilon} = [2.5 ,  5]\times r_c^f$ represented by the blue, red and green shaded bands for \spin{0}, \spin{1} and \spin{2}, respectively. The former exhibits the highest values for this quantity. The black dashed line corresponds to $r_{\epsilon}\sim 3.5r_c^f$, reported in~\citet{Amin_2022}. We observe in both panels of  Fig.~\ref{fig:DMPRofMM}  that the NFW tails for \spin{0} have a wider variation around the mean than for the other spins. This occurs because the scalar field concentrates more mass in the central soliton, leading to a lower probability of occupational density in the outer regions. As a result, the tails become more diverse as the number of solitons changes. This behaviour is less pronounced in the \spin{1} and \spin{2} cases due to smaller interference patterns in the outer regions, leading to a smoother transition between the  core and the tail, which becomes closer to the average profile regardless of the number of initial solitons.

In Fig.~\ref{fig:FitsProf} we compare the spherically averaged halo density obtained directly from the simulations (Fig.~\ref{fig:DMPRofMM}) 
and the fits using \eqref{eq: join_profile}. Without loss of generality, we only show the result for $\Nsol=25$, since the rest of the simulations display a similar behaviour with more (or less) pronounced effects 
which depend on the number of cores in the merger. We can observe a good match for both the core and the tail for simulations and fits.
\begin{figure}
    \centering
    \includegraphics[width=0.9\linewidth]{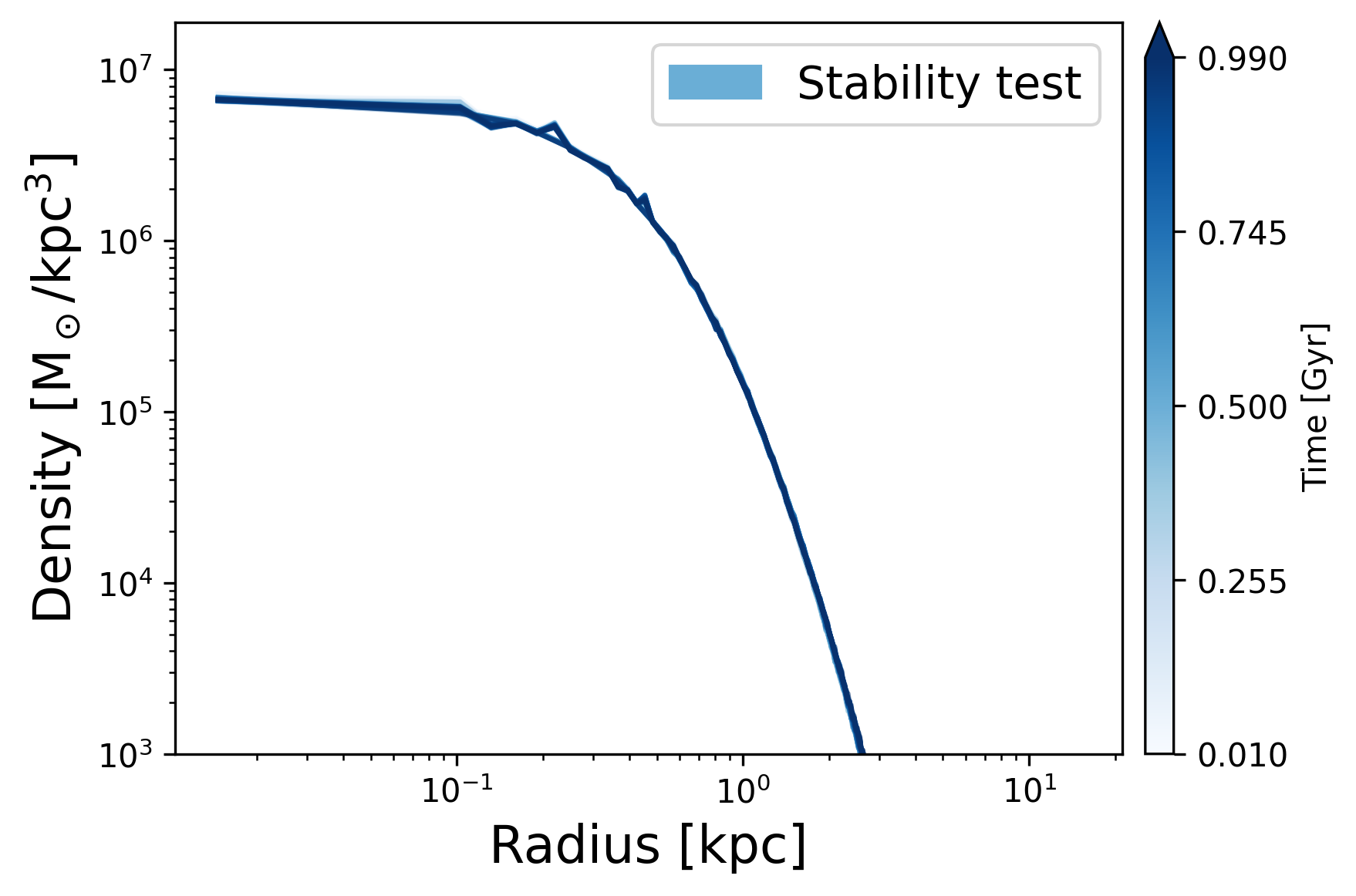}
    \caption{Stability test of the stellar system evolved with our particle-mesh $N$-body code. The system remains stable over 1~Gyr, justifying its use as initial conditions for the study of dynamical heating.}
    \label{fig: stabilityStars}
\end{figure}
\begin{figure*}
    \centering
    \includegraphics[width=0.9\linewidth]{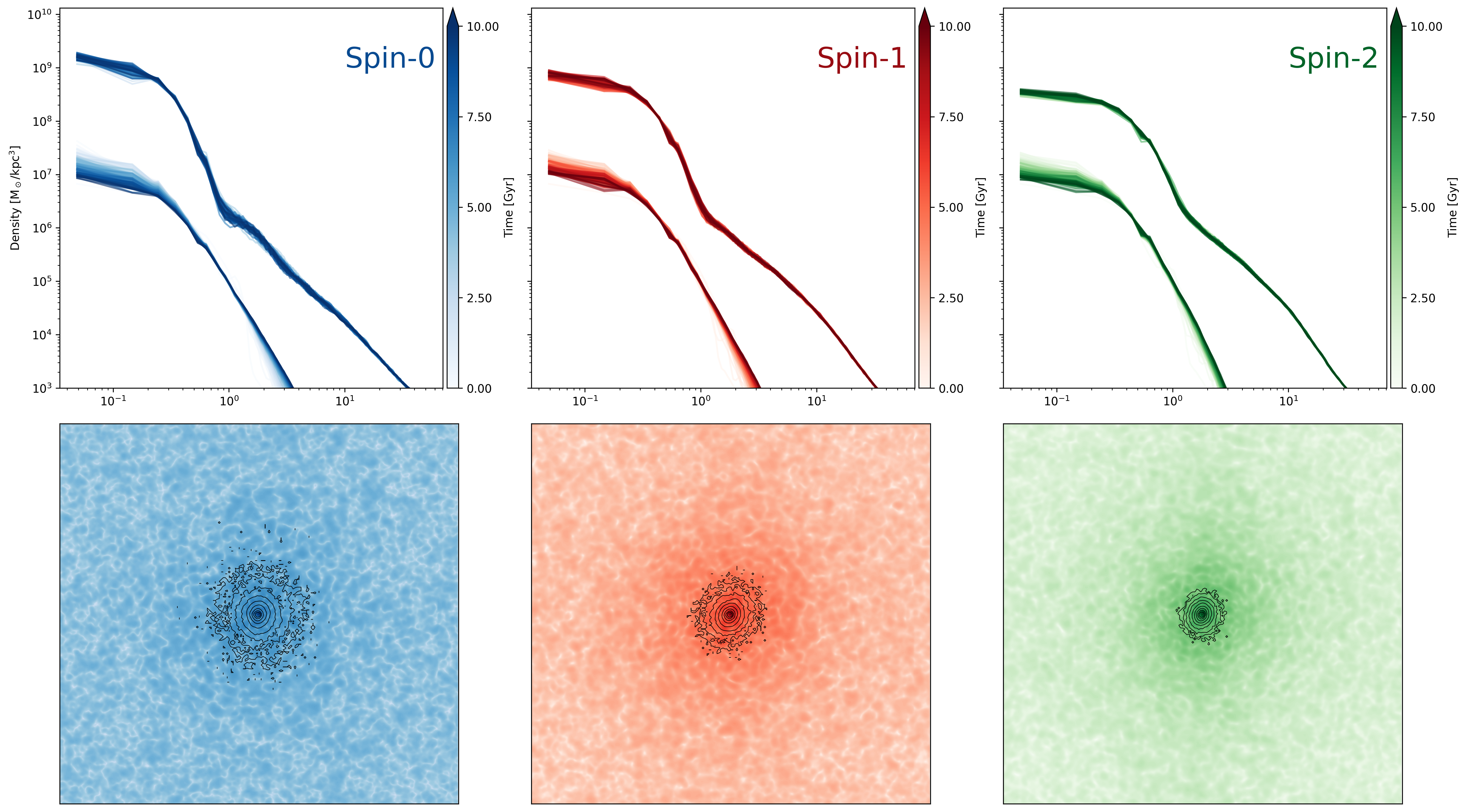}
    \caption{Density distribution from the simulations during 10~Gyrs of evolution. The host halo was constructed merging 25~solitons of equal mass, randomly distributed in all models and taking the resulting wave packet after 5$T_\text{dyn}$. \textbf{Top panels}: From left to right, we show the time evolution of the radial density profile $\rho(r)$ for $10^6$ particles embedded in a host halo composed of \spin{0}, \spin{1} and \spin{2} fields, respectively and the density profile of the host halo. \textbf{Bottom panels}: From left to right, 2D projections of the plane $z = 0$ of the final density distribution of the host halo for \spin{0}, \spin{1} and \spin{2} fields, respectively. The black contours indicate stellar density up to six orders of magnitude below the maximum value at the final time. We can observe that the heating process is spin-dependent.}
    \label{fig: DensityHeat}
\end{figure*}
\subsubsection{Spin scaling relation}\label{sec:SpinAv}

The spin density is defined as in~\citet{Jain:2023ojg}
\begin{equation}\label{eq: spin}
    \boldsymbol{s}_i=i\hbar \epsilon_{ijk}[\Psivec\Psivec^\dagger]_{jk},
\end{equation}
where $[\Psivec\Psivec^\dagger]_{jk} = \Psivec_i \Psivec^\dagger_j$ and $[\Psivec\Psivec^\dagger]_{jk} = \Psivec_{ik} \Psivec^\dagger_{kj}$ 
for \spin{1} and \spin{2} respectively. The spin angular momentum is a \emph{conserved} quantity, obtained as the integral of the spin density over the volume
\begin{equation}\label{eq: spin_vol}
    \boldsymbol{S}_i=i\hbar\int_{\text{vol}}  \epsilon_{ijk}[\Psivec\Psivec^\dagger]_{jk} dV.
\end{equation}
Since $ \boldsymbol{S}_i$ is conserved, its integral over the whole box, $|\boldsymbol S_{\text{tot}}|$, should be the same before and after the merger.

We computed the spin density for \spin{1} and \spin{2} models, finding that in both cases the solitonic core is polarised, that is, $ \boldsymbol{s}_i$ points to a specific direction. However, in the outer regions away from the core, $ \boldsymbol{s}_i$ is randomly oriented from point to point. This agrees with and extends the results of~\citet{Amin_2022}, which focussed on \spin{1}. Fig.~\ref{fig:Spin} shows the relation between the spin density in the solitonic core $|\boldsymbol S_{\text{core}}|$, defined using \eqref{eq: spin_vol} within a spherical volume of radius $2 r_c$, and the total spin $|\boldsymbol S_{\text{tot}}|$ defined from \eqref{eq: spin_vol} over the whole simulation volume. We normalised $|\boldsymbol S_{\text{core}}|$ and $|\boldsymbol S_{\text{tot}}|$ to the total number of particles $\Ncore \equiv M_{\text{core}}/ m_s$ of the core and $\Ntot  \equiv M_{\text{tot}} / m_s$ of the whole simulation box, respectively. We divided the $S_{\text{tot}}/\Ntot$ axis into $40$ equal bins of width $0.025$ each, and determined the average value of $\boldsymbol S_{\text{core}}/\Ncore$ in each bin, depicted by the red and green dots, as well as the standard deviation depicted by the error bars.

We observe a rough correlation between the core $\boldsymbol S_{\text{core}}/ \Ncore$ and the total $\boldsymbol S_{\text{tot}}/\Ntot$ for both cases, with \spin{1} reaching higher values of $|S_{\text{core}}|$ per particle. The initial assignment of spin to solitons is random and due to conservation of $\boldsymbol{S}_i$ this is reflected in the final $\boldsymbol S_{\text{tot}}/\Ntot$. For simulations with larger $N_\text{sol}$ there is more freedom to average out the total spin angular momentum and so those typically correspond to smaller $\boldsymbol S_{\text{tot}}/\Ntot$. The case of \spin{2} has more spin configurations per initial soliton than the case of \spin{1}, resulting in additional compactness in  $\boldsymbol S_{\text{tot}} /\Ntot $. The solid line represents the ideal case where the spin per particle in the core is the same as the total spin and since  \spin{1} has denser cores than \spin{2},
it generally leads to higher $\boldsymbol S_{\text{core}}/\Ncore $ reflecting a higher degree of polarisation of the final soliton.

\section{Dynamical heating}\label{sec: granularities}

\begin{figure}
    \centering
    \includegraphics[width=0.9\linewidth]{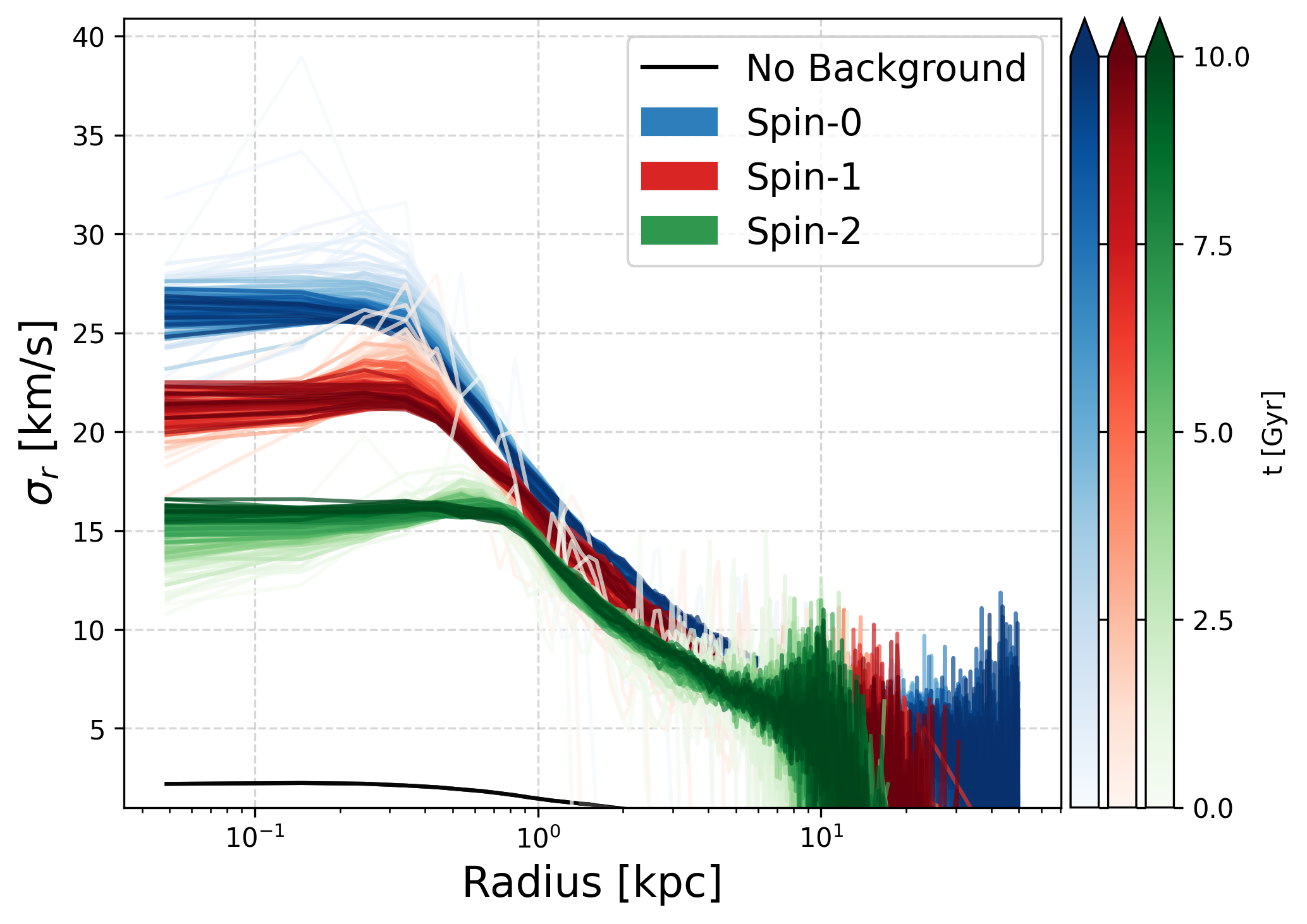}
    \caption{Radial velocity dispersion of the stars. The colour gradient shows the time evolution and the colours represent the effect of different host haloes on stars. They are also ordered from top to botom: blue, red and green for \spin{0}, \spin{1}, and \spin{2} respectively. 
    }
    \label{fig: VelDispHeat}
\end{figure}
The velocity dispersion $\sigma_v$ of a galactic halo provides insights into the study of the substructure and gravitational perturbations produced by dark matter density fluctuations. This phenomenon, known as dynamical heating, has been widely studied for stellar populations of galactic disks. In ULDM models, heating mechanisms can be related to subhalo perturbations or to time-dependent fluctuating substructure due to interference patterns cause by the wave nature of ULDM~\citep{church2019heating}. This has been particularly explored in the case of \spin{0} ULDM, showing that the quantum interference patterns can be an efficient source of heating of galactic disks~\citep{chowdhury2023dynamical,kawai2022analytic,dalal2022excluding}. In higher spin ULDM models, the interference patterns are in general different, a fact which  can then impact the velocity dispersion of the halo.

To investigate the dynamical heating process in spin-$s$ models, we simulate an idealised system of approximately $10^6$ particles, each representing a star with a mass of $3.6 M_\odot$. These particles are embedded in the final halo configuration corresponding to each spin case. Our objective is to trace how the granularities and the mass distribution in the ULDM haloes affect stellar velocities.

The stellar distribution was initialised using the \textsc{GALIC} code \citep{Galic}, adopting a Hernquist profile \citep{1990ApJHernquist}. \textsc{GALIC} is an iterative method for constructing $N$-body galaxy models in collisionless equilibrium. The subsequent evolution of the system was performed using our own Cloud-In-Cell (CIC) $N$-body code. Fig.~\ref{fig: stabilityStars} shows a stability test of the stellar profile after 1~Gyr of evolution, confirming that the stellar configuration is indeed stable.

Once the stability of the system was verified, the stars were embedded within a ULDM halo. These haloes result from the merger of $N_\text{sol} = 25$ solitons after $5\tau_\text{dyn}$. The two components then evolved together during $t_f = 10$~Gyr, meaning that the total potential includes contributions from both the halo and the stars. This combined potential is updated at each time step when solving the Poisson equation for the system.

\begin{figure}
    \centering
    \includegraphics[width=0.9\linewidth]{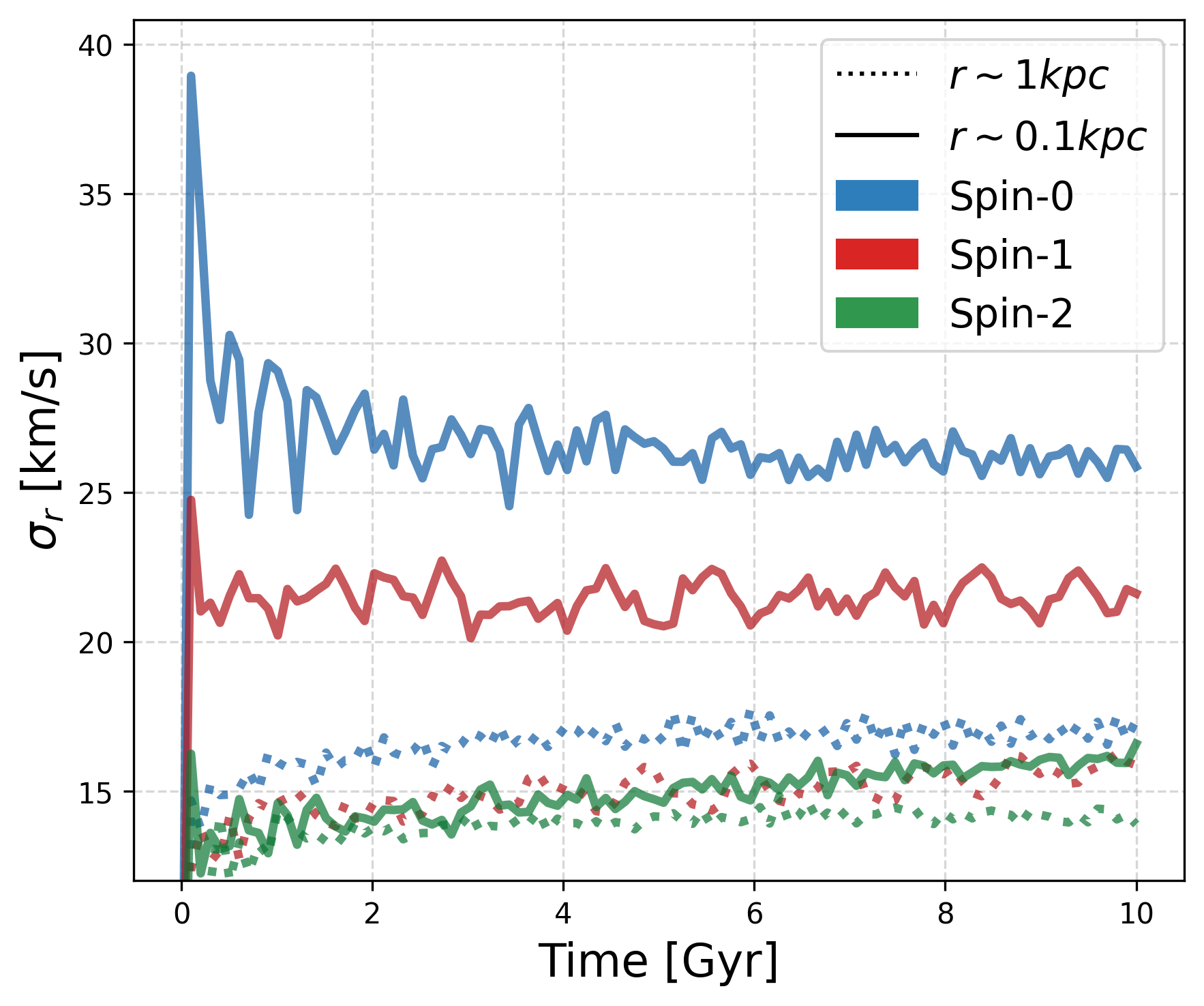}
    \caption{Time dependence of the velocity dispersion of the stars, $\sigma_r$. The colours represent the evolution of different host haloes on stars. They are also ordered from top to botom in each scenario (solid and dashed curves): blue, red, and green for \spin{0}, \spin{1} and \spin{2} respectively. 
    }
    \label{fig: VelDispHeat_Time}
\end{figure}

In Fig.~\ref{fig: DensityHeat}, top panel, we show the evolution of the stellar density alongside that of the corresponding host haloes for spin~0, spin~1 and spin~2, respectively. 
In the bottom panel of the same figure we display the projected density of the host halo on the $z=0$ plane, together with the contours of the stellar orbits at $t = 10$~Gyr. The final size of this system shows how dynamical heating varies between models, even under identical initial conditions, as the interference patterns become less pronounced when the spin is larger~\citep{Amin_2022}. 
In fact, the \spin{0} ULDM halo exhibits the highest density, while the stellar profile becomes more extended over time. This effect is less evident in the \spin{1}and \spin{2} models.

Fig.~\ref{fig: VelDispHeat} shows the evolution of the stellar velocity dispersion profile, $\sigma_v$. To compute $\sigma_v$, the radial coordinate $r$ was divided into bins and the average velocity dispersion was calculated using all the points within each bin. The profiles are plotted with a colour gradient representing the progression over time, where more transparent colours correspond to earlier times and more saturated colours to later times. The black solid line represents the initial isolated condition. We observe that the particles evolve differently for each spin, influenced by constructive and destructive interference as well as variations in core size, resulting in distinct velocity perturbations. The case of spin~0 shows larger $\sigma_v$ at smaller radii, indicating that perturbations are more pronounced in the central regions of the halo, consistent with its denser core compared to the other models (see Fig.~\ref{fig:DMPRofMM}). In contrast, the spin~1 and spin~2 cases display a hierarchical behaviour for $\bar{\sigma}_v$, reflecting the hierarchy observed in their inner core densities.

For better visualisation, Fig.~\ref{fig: VelDispHeat_Time} shows the evolution of $\sigma_v$ over time at two characteristic radii: one near the centre, $r \sim 0.1$~kpc, and the other near the outer edge of the core, $r \sim 1$~kpc, for all ULDM haloes, close to the transition between the core and the NFW tail. Notably, the separation between the curves is the largest for the \spin{0} case, consistent with its overall trend of increased velocity dispersion and more efficient dynamical heating.

\begin{figure}
    \centering
    \includegraphics[width=0.9\linewidth]{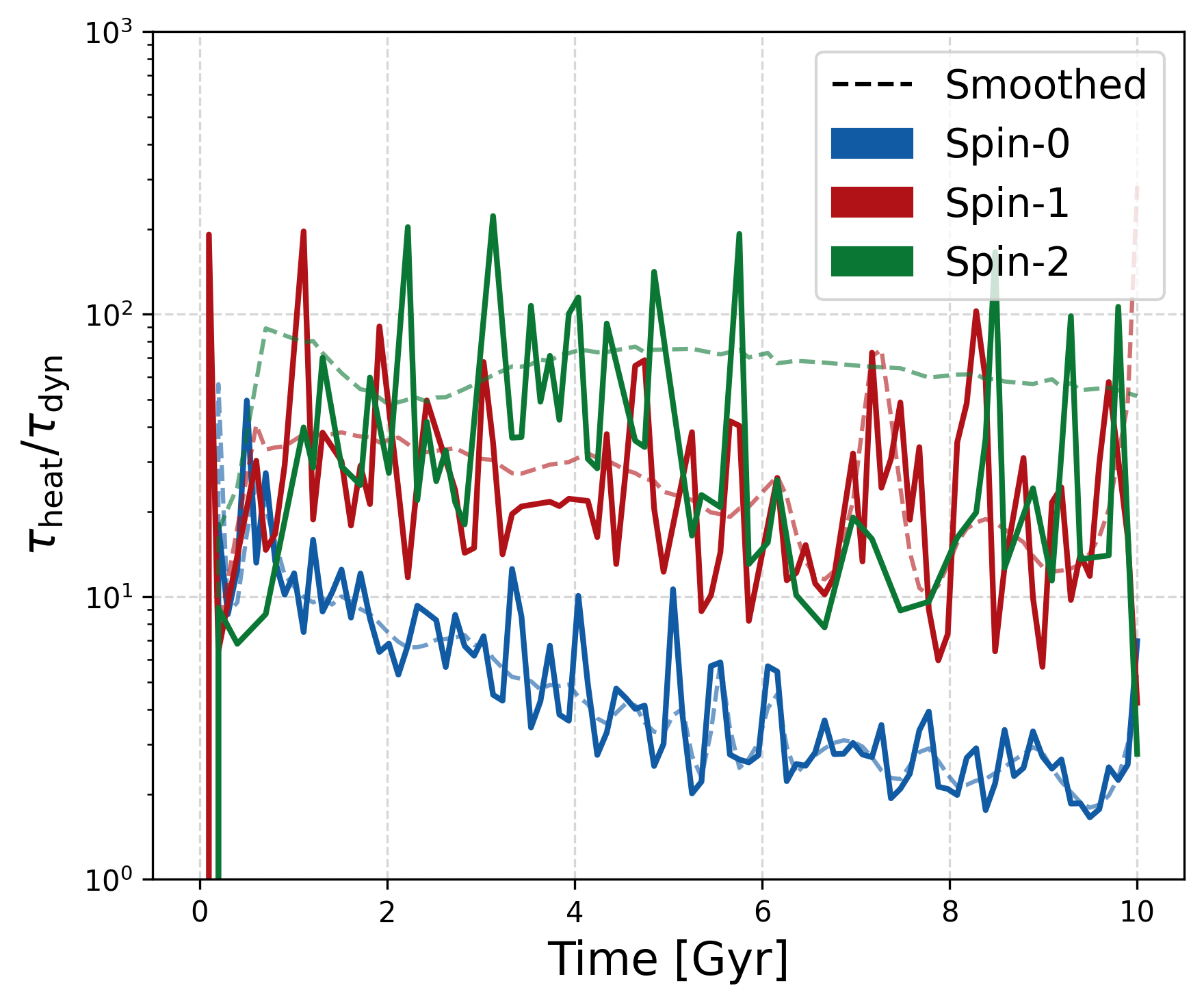}
    \caption{Evolution of the heating time-scale relative to the virialization time-scale over time for the three models. A higher ratio indicates lower heating efficiency. Notably, haloes with higher spin exhibit significantly larger values of this ratio throughout their evolution, implying that they remain as quasi-equilibrium over longer periods. The dashed lines represent smoothed profiles.}
    \label{fig:heating}
\end{figure}

Following a similar analysis as in \citet{chowdhury2023dynamical}, we show the ratio between the time evolution of the
dynamical heating time $\tau_{\text{heat}}=\displaystyle\frac{r_{\text{gal}}}{dr_{\text{gal}}/dt}$ and the dynamical time in Fig.~\ref{fig:heating}.
We observe that this ratio exceeds unity across all models, indicating that the system reconfigures dynamically faster than it accumulates net heating. This suggests that dynamical heating is inefficient, as the injected energy is spread over multiple dynamical timescales. Moreover, this effect is amplified by up to an order of magnitude for higher spin, indicating that the system can be described as a quasi-equilibrium configuration for a longer duration. Even within the limitations of our simulations, this provides valuable insight into how spin-$s$ ULDM can help maintain stellar configurations in equilibrium, potentially explaining the long survival times of globular clusters, such as those observed in Fornax. Specifically, the formation of more extended and less centrally concentrated structures reduces the effects of dynamical friction. Moreover, the suppression of dynamical heating may allow stellar clusters to persist without being rapidly drawn toward the galactic centre.

\section{Scaling relations for density profiles of ULDM haloes}\label{sec: URSDP}

The SP system allows the rescaling of soliton solutions $\{M,m_s\}\to  \{\lambda M, \beta m_s\}$ as
\begin{equation}
    \left\{t, x , \psi, \rho\right\} \to \left\{\lambda^{-2}\beta^{-3}t, \lambda^{-1}\beta^{-2}x , \lambda^{2}\beta^{3}\psi,\lambda^{4} \beta^{6} \rho\right\},
\end{equation}
leaving the system unchanged, see \eqref{eq:scaling} in Appendix~\ref{a: GS-soliton}. This leads to scaling relations which we investigate in this section, particularly their time dependence as the system relaxes towards equilibrium. For this, we use the same set of $24$ simulations as in section~\ref{sec: MBMergers}. Our aim is to be able to infer the final state of the merger of an initial number of solitons, $\Nsol $, using such scaling relations.

\subsection{Scaling relations for central solitons} \label{subsec: SRFCC}

As discussed in Appendix \ref{a: GS-soliton}, the scaling symmetries of the SP system allow the rescaling of soliton solutions, in particular, a single soliton mass follows the relation $M_{\text{sol}}\rightarrow \lambda M_{\text{sol}}$. \citet{Amin_2022} argue of a relation  $M_{\text{sol}}^f \propto \Nsol  M_{\text{sol}}^i$ and demonstrate a tight correlation between $M_{\text{core}} /M_{\text{tot}}$ and a measure of the total energy of the system. We take this idea further and investigate the existence of similar relations between the characteristic parameters $r_c^i$ and $\rho_c^i$ that describe the initial solitons with $r_c^f$ and $\rho_c^f$ that describe the final ULDM core, and their dependency on time until asymptotic relaxation. We focus on $\lambda_{\rho}$ given by
\begin{equation}
\label{eq:lambdas}
    \lambda_{\rho}=\left(\frac{\rho_c^f}{\rho_c^i}\right)^{1/4},
\end{equation}
which contains information about the halo's characteristic maximum density.

\begin{figure}
    \centering
    \includegraphics[width=0.9\linewidth]{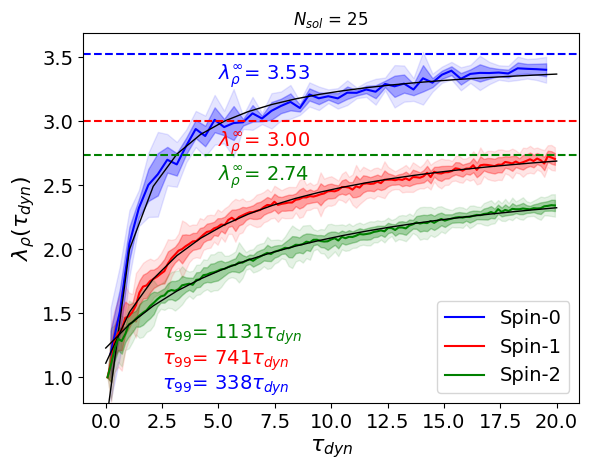}
    \caption{
 Time evolution of $\lambda_\rho$ for $\Nsol=25$  until $t=20\tau_{\text{dyn}}$. Black curves show the best fit according to \eqref{eq:satFun}. The curves are ordered from top to botom for \spin{0}, \spin{1}, and \spin{2}, respectively. The horizontal dashed lines mark the  $\lambda^{\infty}_{\rho}$ saturation limit, assuming convergence. The value for $\tau_{99}$, denoting the number of dynamical times required to achieve $99\%$ of the value of $\lambda^{\infty}_{\rho}$ is also displayed.}
    \label{fig: lambda_inf}
\end{figure}

As mergers undergo a relaxation process before forming the final halo, $\lambda_{\rho}$ will evolve until it reaches a saturation value which is  when the system is fully stabilised. We calculate $\lambda_{\rho}$ from our simulations by tracing the maximum density in the box to define $\rho_c^f(t)$ at time $t$ and use it in \eqref{eq:lambdas} along with $\rho_c^i$. We do this for all our 24 simulations indexed by $\Nsol $ and the spin. Fig.~\ref{fig: lambda_inf} displays the evolution of $\lambda_{\rho}$ as a function of $t/\tau_{\text{dyn}}$ for the case $\Nsol =25$. The solid lines (blue for \spin{0}, red for \spin{1} and green for \spin{2}) indicate the smoothed mean value of the $\lambda_\rho$ parameter from the simulations. The dark and light-shaded regions represent $1\sigma$ and $2\sigma$ deviations, respectively. The black solid curve corresponds to the best fit using the  saturation function
\begin{equation}\label{eq:satFun}
  \lambda_{\rho}(t)  = a + \frac{b t}{1+c t}.
\end{equation}
The saturation value $\lambda^{\infty}_{\rho}$, displayed in the figure, is marked for each spin case as horizontal dashed lines. We also show  $\tau_{99}$, defined as the value of $t/\tau_{\text{dyn}}$ where $\lambda_{\rho}$ reaches 99$\%$ of its saturation value. Appendix \ref{sec: lambda_conv} displays the evolution of $\lambda_{\rho}$ for different spatial resolutions concerning the \spin{0} model. This is a consistency test to complement the discussion shown in Appendix \ref{sec: stability}. 

The \spin{0} case has more interference patterns resulting in a higher value for $\lambda^{\infty}_{\rho}$, consistent with having higher central density as in Fig.~\ref{fig:DMPRofMM}. This leads to a larger saturation value $\lambda^{\infty}_{\rho}$ with a steeper initial slope to reach it. This is less so for \spin{1} and even less so for the \spin{2} case. We also observe that the relaxation time $\tau_{99}$ shows a hierarchical behaviour, wherein the \spin{0} case has the largest value, followed by the \spin{1} and then \spin{2} cases. This is verified for any number of initial solitons, not only for $\Nsol =25$, see Appendix~\ref{sec: lambda_infty}.

Additionally, we observe a monotonic increasing relation between $\lambda_{\rho}(t)$  and $\Nsol $ as a function of the dynamical time $\tau_{\text{dyn}}$ when the densities $\rho_c^f$ and $\rho_c^i$ were computed. In fact, if the system evolves over longer dynamical times, the mergers with higher $\Nsol$ will result in higher values of $\lambda_{\rho}^{\tau_{\text{dyn}}}\equiv \lambda_{\rho}(t=\tau_{\text{dyn}})$; moreover, the larger the dynamical time, the larger the density ratio $\lambda_{\rho}^{\tau_{\text{dyn}}}$. In Appendix \ref{sec: lambda_infty}, we analyse the behaviour of $\lambda_{\rho}^{\tau_{\text{dyn}}}$ as a function of $\Nsol$ across various dynamical times. We observe a clear trend of convergence toward the asymptotic curve $\lambda_{\rho}^{\infty}$. For simplicity, we will focus on $\lambda^{20}_{\rho}$, in the subsequent sections.

\begin{figure}
    \centering
    \includegraphics[width=0.8\linewidth]{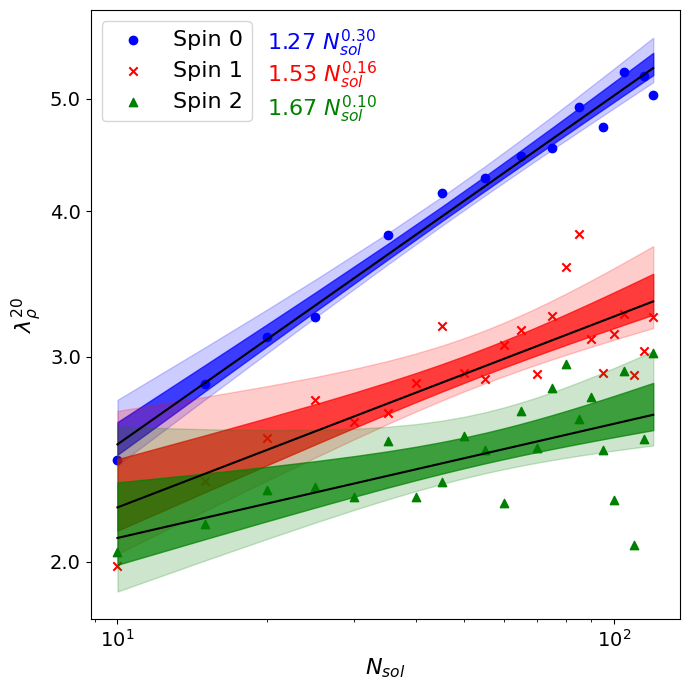}
    \caption{The scaling of $\lambda^{20}_{\rho}$ with $\Nsol$ for all simulations in each spin~$s$ model. The solid line corresponds to the best fit, and the dark-shaded (light-shaded) band represents the $1\sigma$ ($2\sigma$) standard deviation away from the best fit. The best-fitting relations are also depicted in the figure.}
    \label{fig: lambda_bar}
\end{figure}

We found that the relation of $\lambda^{20}_{\rho}$  as a function of the number of initial solitons $\Nsol$ can be fit by the following power law
\begin{equation} \label{eq: sqrt_fit}
    \lambda_{\rho}^{20}(\Nsol)= A_{\lambda}\Nsol^{B_{\lambda}}
\end{equation}
with $A_{\lambda}=\{1.27, 1.53, 1.67\}$ and $B_{\lambda}=\{0.30, 0.16, 0.10\}$ for \spin{0}, \spin{1} and \spin{2}, respectively. This is displayed in Fig.~\ref{fig: lambda_bar} for all three spin models. The blue points, red crosses, and green triangles represent the value of $\lambda^{20}_{\rho}$ for \spin{0},  \spin{1} and \spin{2}, for each simulation indexed by $\Nsol$. The black lines represent the best fit for each model with the fitting function displayed on the figure, assuming $\lambda_\rho$ depends only on $\Nsol $. The dark and light-shaded regions represent the 1$\sigma$ and 2$\sigma$ standard deviations away from the best fit. We observe a hierarchy in the slope of the best fit between the spins, with \spin{0} being the steepest. This is consistent with our findings of section~\ref{sec: dens} which indicates that  the lower the spin, the more compact haloes  with higher central densities form by the merger of the same number of solitons.

Given the relation for $\lambda^{20}_{\rho}$ just found  and that $N_\text{sol}$ is related to the total mass of the system, we may determine a scaling relation between the initial and final mass of the soliton configurations as
\begin{equation}
    M_c^f=\lambda^{20}_{\rho}(\Nsol)M_c^i.
\end{equation}
This has the advantage being able to  characterise the resulting soliton after several $\tau_\text{dyn}$ of evolution without the need to run the simulations.
\subsection{Scaling relations for the NFW-tail}\label{sunsec: NFW-Fit}

The outer regions of haloes are characterised by an NFW tail described in~\eqref{eq: nfw}. We explored the evolution of $r_{\epsilon}$ and $r_s$ as a function of $\Nsol$, evaluated at $t = 20 \tau_{\text{dyn}}$,  using the simulations of section~\ref{sec: dens}. For this analysis, we fit first the core of the halo and then use the corresponding $r_c^f(\tau_{\text{dyn}}=20)$ value to normalise the parameters that characterise the tail, $r_s$ and $r_\epsilon$.

\begin{figure}
    \centering
    \includegraphics[width=0.76\linewidth]{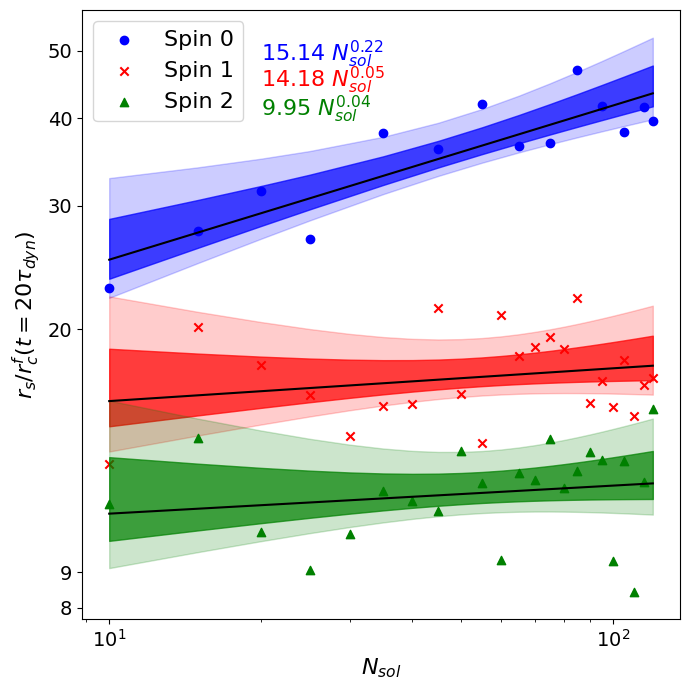}
 \caption{The scaling of $r_s/r_c^f$ with $\Nsol$ for all simulations in each spin~$s$ model, computed at $t = 20\tau_{\text{dyn}}$. 
The black line corresponds to the best fit,
and the dark-shaded (light-shaded) band represents the $1\sigma$ ($2\sigma$) standard deviation away from the best fit. The best-fitting relations are also shown.}     
    \label{fig: rsol_Nsol}
\end{figure}

We find that $r_s/r_c^f$ is well fitted with the same functional form as in \eqref{eq: sqrt_fit}, that is, $ r_s/r_c^f = A_{s}\Nsol^{B_{s}}$, with the following values $A_{s}=\{15.14, 14.18, 9.9\}$ and $B_{s}=\{0.22, 0.05, 0.04\}$ for \spin{0}, \spin{1} and \spin{2}, respectively. The result is graphically displayed in Fig.~\ref{fig: rsol_Nsol}, where we see that \spin{0} requires a significantly higher value for $r_s$ than \spin{1}, which is marginally higher than the \spin{2} model. This implies that mergers with the same number of solitons result in less steep tails for \spin{0} compared to either \spin{1} or \spin{2}.

\begin{figure}
    \centering
    \includegraphics[width=0.76\linewidth]{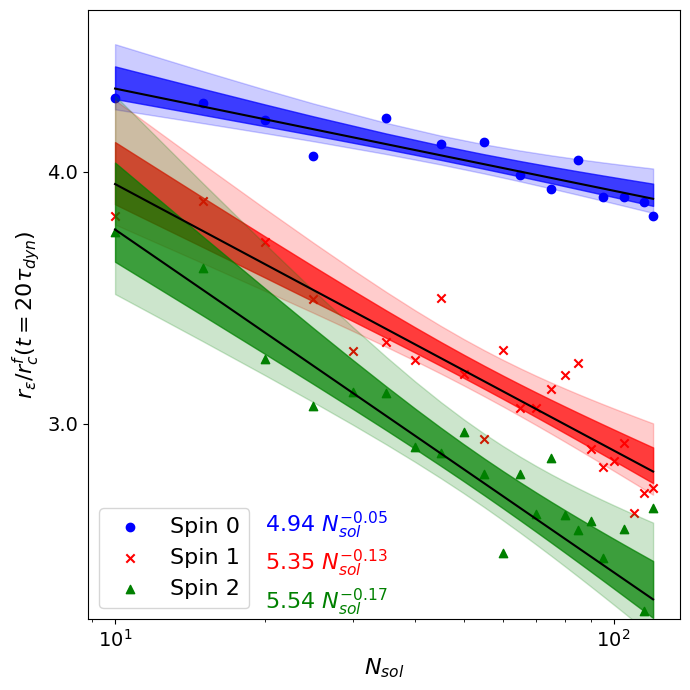}
 \caption{The scaling of $r_{\epsilon}/r_c^f$with $\Nsol$ for all simulations in each spin~$s$ model, computed at $t = 20\tau_{\text{dyn}}$. 
The black line corresponds to the best fit,
and the dark-shaded (light-shaded) band represents the $1\sigma$ ($2\sigma$) standard deviation away from the best fit. The best-fitting relations are also shown.}     
    \label{fig: reps_Nsol}
\end{figure}

Finally, $r_{\epsilon}/r_c^f$ is once more fitted with the same functional form as in~\eqref{eq: sqrt_fit}, that is, $ r_{\epsilon}/r_c^f = A_{\epsilon}\Nsol^{B_{\epsilon}}$, with the fitting parameters taking values in $A_\epsilon = \{4.94, 5.35, 5.54\}$ and $B_\epsilon = \{-0.05, -0.013, -0.17\}$  for \spin{0}, \spin{1} and \spin{2} respectively;  see Fig.~\ref{fig: reps_Nsol}. We see that $r_{\epsilon}$ is larger in the \spin{0} model, followed by \spin{1} and \spin{2} respectively, meaning that  \spin{0} transitions more slowly from the solitonic core to the NFW tail.

The main conclusion from Figs.~\ref{fig: lambda_bar}, \ref{fig: rsol_Nsol} and \ref{fig: reps_Nsol} is that \spin{0} configurations produce more compact solitons with higher central densities. These solitons enclose more mass than \spin{1} and \spin{2} configurations, as the transition from the soliton to the NFW tail occurs at larger radii in \spin{0} models. In fact, the lines for $\lambda_{\rho}^{20}$, $r_s$, and $r_{\epsilon}$ do not intersect for positive values of $\Nsol$, indicating that this behaviour remains consistent regardless of the number of initial configurations. Specifically, each model has distinct regions for $r_{\epsilon}$, as shown by the shaded bands in Fig.~\ref{fig:DMPRofMM}. This suggests that the density profiles for each model display distinct characteristics that can be contrasted with observations.

Using this information, the density profile of each model can be characterised by the halo's central density and the number of initial soliton configurations, which can even be non-integer values. In this sense, we can create equivalent haloes with the same mass and corresponding density profile for each spin~$s$ model. This will be applied in the following section, using an equivalent host halo for \spin{0}, \spin{1} and \spin{2} configurations. 

It is important to point out that in more realistic scenarios, such as those based on cosmological simulations, or when including baryonic effects, the shape of the NFW tail will be modified. 
In this sense, our analysis illustrates how varying the ULDM spin affects not only the core but the entire dark matter profile, even under identical initial conditions. 
These differences motivate further studies aiming to compare theoretical predictions with observations. Lastly, while we expect that the differences between spins may be less pronounced in more realistic scenario, the inclusion of the ULDM self-interactions will instead enhance them; we leave this possibility for future work.

\section{Soliton cores as satellite haloes}\label{sec: Sathaloes}

In this section we investigate how the ULDM spin influences the tidal disruption of a satellite orbiting a ULDM halo. To simplify the analysis, we treat the host halo as an external potential, following a similar process as in \citet{du2018tidal}. For each model, we construct equivalent haloes by applying the scaling relations introduced in Sec.~\ref{sec: URSDP}, ensuring that either the core size or the total halo mass are kept fixed across the models. This method allows us to define the complete ULDM density profile. We will show that, for satellites orbiting in the outer regions of these haloes (beyond the core), their dynamics is primarily governed by the tail of the halo. In fact, the structure of this tail is determined by scaling relations that are closely linked to the spin-dependent properties of the system.

\subsection{Constructing the system: host halo and satellite}

\begin{figure}
    \centering
    \includegraphics[width=0.8\linewidth]{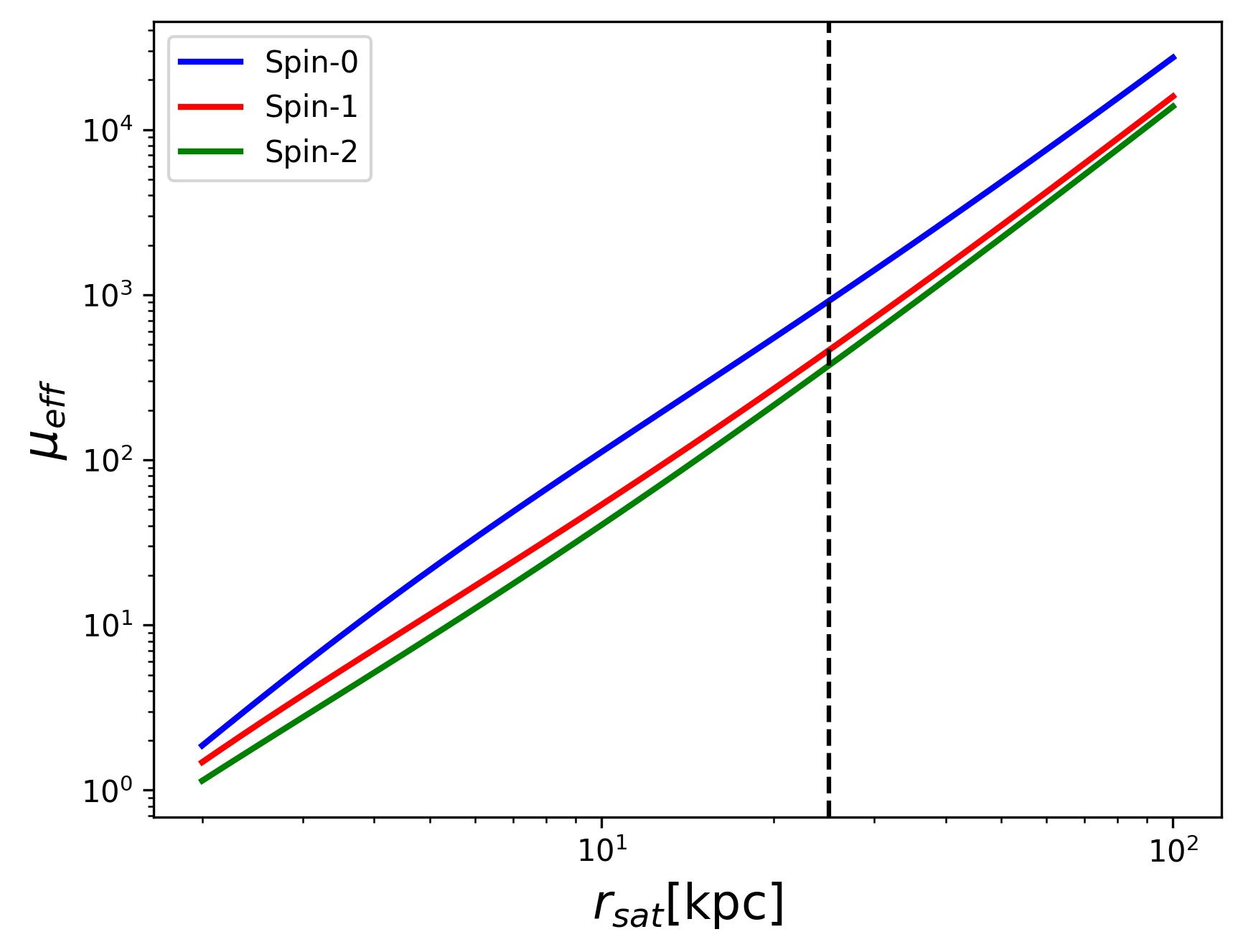}
    \caption{Relation between $\mu_{\text{eff}}$ and $r_{\text{sat}}$ when the scaling radius is fixed to $r_c=0.15$ kpc, which implies that the mass of the scalar field has to be modified in each case by a factor of $\beta=\{1.03,1.25,1.38\}$. We show the dashed vertical line at $r_{\text{sat}}=25$ kpc to compare the different values of $\mu_{\text{eff}}$ for each model at the position of the satellite. }
    \label{fig:case1}
\end{figure}


\begin{figure}
    \centering
    \includegraphics[width=0.8\linewidth]{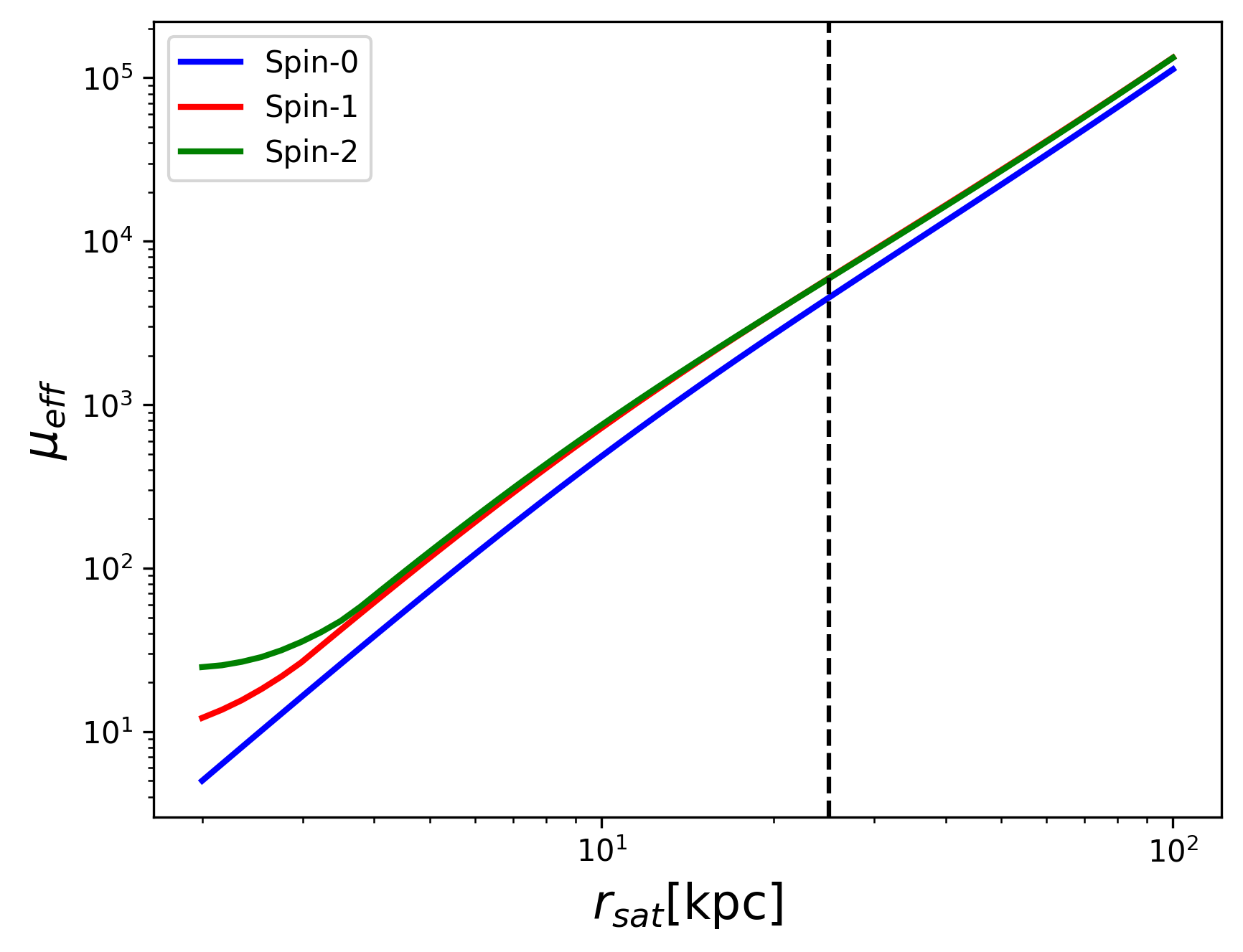}
    \caption{Relation between $\mu_{\text{eff}}$ and $r_{\text{sat}}$ when $M_{200}=1\times 10^9 M_{\odot}$ is fixed for all models for the same range as in Fig.~\ref{fig:case1}. In this case the haloes must be rescaled by a factor of $\lambda=\{0.38, 0.30, 0.29\}$. The dashed black line represent the position of the satellite.}
    \label{fig:case2}
\end{figure}

\begin{figure}
    \centering
    \includegraphics[width=0.8\linewidth]{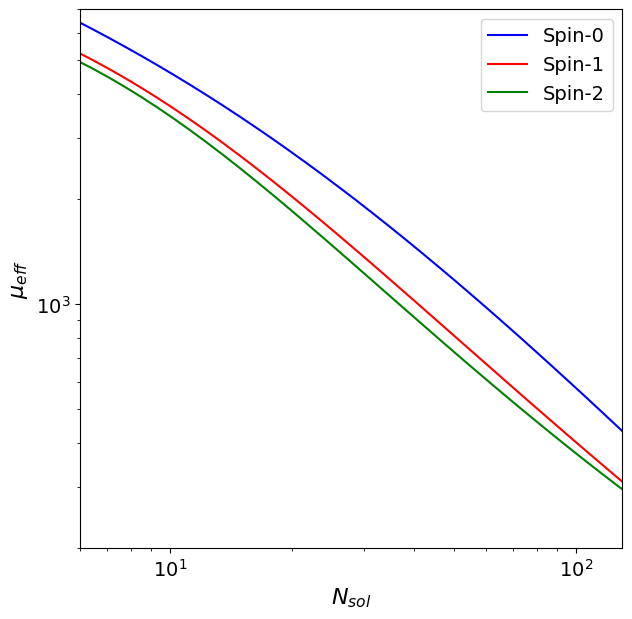}
    \caption{The dimensionless effective density parameter $\mu_{\text{eff}}$ as a function of the number of solitons $\Nsol$ for each spin~$s$ model. In this case, $\rho_{c,0}^{\text{sat}}=1.37\times 10^7M_{\odot}/\text{kpc}^3$ and $r_\text{sat}=25$ kpc are fixed.  } 
    \label{fig:mu_Nsol}
\end{figure}

In order to consider the ULDM halo as a host of a satellite configuration, we will define the effective mean density of the host halo as   
$
    \bar{\rho}_{\text{eff}}= \bar{\rho}_{\text{halo}}(r_{\text{sat}}) - \rho_{\text{halo}}(r_{\text{sat}}),
$
where $\bar{\rho}_{\text{halo}}(r_{\text{sat}})$ is the average density of the halo computed until $r_{\text{sat}}$ and use this to define the dimensionless effective density parameter $
    \mu_{\text{eff}}\equiv \displaystyle\frac{\rho_{c,0}^{\text{sat}}}{\bar{\rho}_{\text{eff}}}
\label{mu_eff}
$, where $\rho_{c,0}^{\text{sat}}$ is the initial central density of the satellite.
We constructed the host haloes such that $\mu_{\text{eff}}$ is the same for each spin, trading  $\Nsol$ with $\mu_{\text{eff}}$ since for a specific value of $\Nsol$, and given the spin, this completely fixes the halo profile for fixed $m_s$ and $M_{\text{sol}}$. 
Fig.~\ref{fig:mu_Nsol} shows the relation between $\mu_{\text{eff}}$ and $\Nsol$. We observe a hierarchical behaviour across the three cases, with \spin{0} exhibiting the highest value. Additionally, since the relations shown in Sec.~\ref{sec: URSDP} are hierarchical over the simulated domain, we expect that this behaviour remains the same as $\Nsol$ increases, leading to a hierarchical behaviour for $\mu_{\text{eff}}$ as well. 

We will consider variations in the total mass of the halo and the fundamental mass of the ULDM theories, $m_s$. This allows us to analyse the dynamics of the satellite considering equivalent systems in terms of a given parameter. We will refer to the mass of the system as $M_{200}=M(r<r_{200})$ where $r_{200}$ is the radius at which the halo's density is 200$\rho_{\text{crit}}$, with $\rho_{\text{crit}}=127.05 M_{\odot}/\text{kpc}^3$.

\subsubsection{Case 1: Same core size}
Our purpose here is to examine ULDM haloes whose core has the same radius $r_c$ in all cases. Although the central regions of galaxies are not yet well characterised by observations, some surveys aim to obtain more accurate measurements \citet{hunter2012little}. In order to keep the same value of $r_c=0.15$ kpc in all models, we considered the scaling relation $m_{\text{ULDM}}\rightarrow \beta m_{\text{ULDM}}$ (see Appendix~\ref{a: GS-soliton}). The mass of the system is $M=\{2.60, 3.29, 3.46\}\times 10^9 M_{\odot}$ and $\beta=\{1.03, 1.25, 1.38\}$ for \spin{0}, \spin{1} and \spin{2}, respectively. In Fig.~\ref{fig:case1}, the relation between $\mu_{\text{eff}}$ and the distance to the satellite $r_{\text{sat}}$ is shown. We observe that $\mu_{\text{eff}}$ and thus the tidal disruption time follows a hierarchical behaviour, being \spin{0} the model with the largest value. Moreover, \spin{1} and \spin{2} show more similar values of $\mu_{\text{eff}}$ over the entire domain.  Because the survival time grows for larger $\mu_{\text{eff}}$ as $\tau \sim e^{\mu_{\text{eff}}}$ \citep{hui2017ultralight}, this implies that a satellite will survive longer when orbiting a \spin{0} ULDM halo compared to haloes with the same core size but composed of ULDM with a higher spin.

\subsubsection{Case 2: Same $M_{200}$}
In this case we instead rescale $M_{200}$ of the ULDM halo so that it is the same across all models. This approach can be applied to astrophysical systems, where the halo mass has been inferred by considering precise measurements of the galactic components. The scaling relation to consider is $M_{200}\rightarrow \lambda M_{200}$, with $\lambda=\{0.38, 0.30, 0.29\}$ for \spin{0},~1 and~2, respectively. The initial mass for each halo are $M_{200}=\{2.60, 3.29, 3.46\} \times 10^9 M_{\odot}$, giving a mass of $M=1\times 10^9 M_{\odot}$ after the transformation. Fig.~\ref{fig:case2} shows the relation between $\mu_{\text{eff}}$ and the distance to the satellite within the same range as in case 1). Here, we observe an inverted hierarchy where $\mu_{\text{eff}}$ become more similar as $r_{\text{sat}}$ increases. In this case, \spin{2} shows the highest value of the tidal disruption time.

\subsubsection{Comparison with previous results}

As a consistency test, we performed three-dimensional simulations by considering the host ULDM halo as an external potential calculated from the fitted ULDM profile of Eq.~\eqref{eq: join_profile}. Similar problems have been studied analytically in~\citet{hui2017ultralight} for the \spin{0} case using a simplified quadratic external potential with spherical symmetry. This was further  explored through wave simulations in~\citet{du2018tidal} by assuming that the external potential is that of a uniform sphere with mass $M_{\text{halo}}$ rather than a realistic ULDM profile that we use here.
In this case, we make the additional assumption that the satellite is in a state of extreme polarisation by setting, without loss of generality, $c_\pol = \delta_{0\pol}$ in Eq.~\eqref{eq:polarisedwf}. This amounts to having the satellite being described by a \spin{0} ULDM soliton and this approximation is valid as long as the satellite remains isolated. Notice that in our actual simulations the satellite is consistently built from spin-$s$ ULDM. As expected, the results were identical as long as the host halo is an external potential.

\begin{figure}
    \centering
    \includegraphics[width=0.85\linewidth]{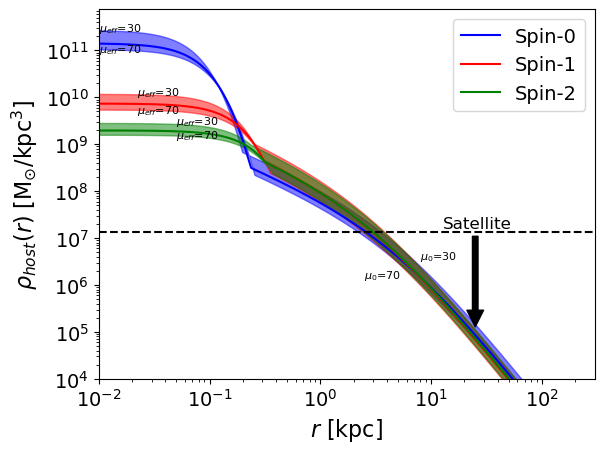}
    \caption{Density profile of the host halo reconstructed given a value of $\mu_{\text{eff}}$ in the range $30$ to $70$.
 The upper boundary of each shaded band represents $\mu_{\text{eff}}=30$ while the lower boundary refers to $\mu_{\text{eff}}=70$. 
In all cases, the initial central density of the satellite is $\rho_{c,0}^{\text{sat}}=1.37\times 10^7M_{\odot}\text{kpc}^{-3}$.
 The arrow indicates the satellite's position relative to the centre of the halo and the horizontal back dashed line shows the value of $\rho_{c,0}^{\text{sat}}$.}
    \label{fig:mu_dens}
\end{figure}

In Fig.~\ref{fig:mu_dens} we show the density profiles of the host haloes, $\rho_{\text{host}}$, for $\mu_{\text{eff}}$ in the range $[30,70]$, reconstructed using the scaling relations from Figures \ref{fig: lambda_bar}, \ref{fig: reps_Nsol} and \ref{fig: rsol_Nsol}. The lowest boundary of the shaded band represents $\mu_{\text{eff}}=70$, the solid line at the centre corresponds to $\mu_{\text{eff}}=50$ and the upper boundary refers to $\mu_{\text{eff}}=30$. Recall that the density profile for \spin{0} exhibits the most pronounced transition, more closely resembling the density of a uniform sphere, characterised by a step function with an average density $\bar{\rho}_{\text{eff}}$ for $r\leq r_*$, where $r_*$ is the radius of the sphere.


\begin{figure}
    \centering
    \includegraphics[width=0.9\linewidth]{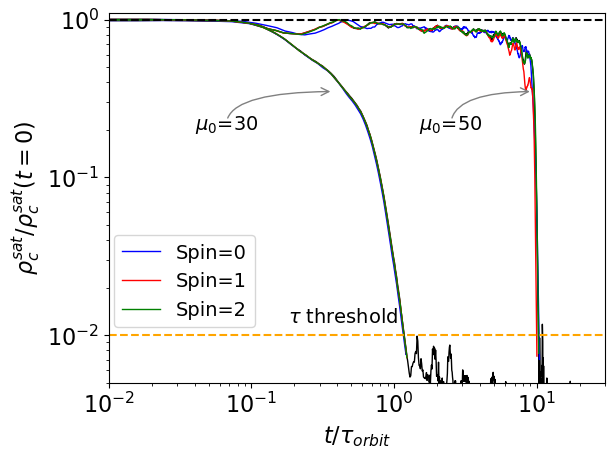}
    \caption{Evolution of the central density normalised to the initial value for $\mu_0 = 30$ and $\mu_0 = 50$ as a function of the number of orbits for  \spin{0} (blue), \spin{1} (red) and \spin{2} (green)
for $\mu_\text{eff} = 30$ and $\mu_\text{eff} = 50$. The case of uniform sphere is shown in black.
 The black dashed horizontal line shows the difference in the initial value of the density. The yellow horizontal line shows the threshold we used to estimate the parameter $\tau$ as in~\citet{hui2017ultralight}.}
    \label{fig: SurvivalTime}
\end{figure}


Fig.~\ref{fig: SurvivalTime} shows the evolution of $\rho_c^{\text{sat}}$, normalised by its initial value, as a function of the number of orbits. All models reproduce the same behaviour as the uniform sphere for a given $\mu_{\text{eff}}$, consistent with the analytical prediction by \citet{hui2017ultralight}, who reported a monotonically increasing relation between $\mu_{\text{eff}}$ and the disruption time $\tau \sim e^{\mu_{\text{eff}}}$. This trend was also confirmed by \citet{du2018tidal} using 3D simulations of uniform spheres. Our results demonstrate that spin-$s$ ULDM models reproduce this behaviour when the host halo is modelled as an external potential, neglecting both granularity and host-satellite interactions. This is because most of the halo mass lies in the outskirts and core effects are negligible.\footnote{This is consistent with \citet{du2018tidal}, which showed that a NFW halo gives the same result as a uniform sphere.}

\section{Conclusion}\label{sec: Conclu}

We performed idealised numerical simulations for the \spin{0}, \spin{1} and \spin{2} ULDM models, finding important differences between them for virialised systems. First, the resulting haloes from merging multiple solitons exhibit notable variations in the density profiles. The \spin{0} model always produces denser, more compact cores with a more prominent transition between the soliton and the NFW-tails. In contrast, the haloes formed in the \spin{1} and \spin{2} models share more similarities, featuring less dense central cores and less extended envelopes with smoother transitions. This is attributed to interference effects, as higher spin values reduce the probability of having fully constructive or destructive interference. In fact, these similarities persist across all the scaling relations observed for the density profile parameters: \spin{0} differs consistently significantly from \spin{1} and \spin{2}. The general shape of the haloes remains consistent regardless of the number of solitons involved in the merger. While more detailed simulations are needed to draw precise conclusions, the distinct features of the spin~$s$ models already offer a basis to distinguish between them, making them a valuable framework for comparison with observational data.
Additionally, notice that in this work we have limited ourselves ULDM without self-interactions. We expect that the introduction of the self-interactions will make the differences between spins even more prominent, because different polarisations in the SP system \eqref{eq:multipleSP} will couple to each other directly, rather than simply through the common gravitational potential $\Phi$. We leave this possibility for future work.

The resulting haloes were used as a host of a stellar system with a Hernquist profile in order to study the dynamical heating process. We found that the velocity dispersion decreases for larger spin since the central density for \spin{0} is higher than \spin{1} and \spin{2}, and the interference patterns are fewer for the latter two models. This result can give an insight of the possibility to relax the constraints on the mass of the ULDM candidates arising from the dynamical heating of stellar systems. Indeed, for the \spin{0} case, it has been argued that for masses below $\sim 10^{-19}$ the dynamical heating would increase the velocity dispersion in ultralight dwarf galaxies to values much larger than what is observed~\citep{chowdhury2023dynamical,dalal2022excluding}. 
We found that the dynamical heating process is up to an order of magnitude less efficient for spin~2 ULDM compared to the spin~0 case. This implies that the system can remain closer to an equilibrium configuration for a longer duration, enabling the stellar system to better preserve its structure. Despite its simplicity, this approach can still provide valuable insight into related phenomena, such as the so-called timing problem observed in some dwarf satellite galaxies, like Fornax. In standard CDM models, a massive dark matter halo would cause globular clusters to experience strong dynamical friction, leading them to lose energy, spiral inward, and eventually merge with the centre of the galaxy. This results in survival times that are much shorter than the observed age of the galaxy. In contrast, in ULDM models, dynamical friction is greatly suppressed. As a result, satellite systems can survive many more orbits before merging, their evolution being primarily determined by the timescale for tidal disruption \citep{hui2017ultralight}. This effect may be even more pronounced when considering higher-spin models.

We emphasise that our results are based on idealised host haloes formed through mergers in simulations. While more detailed modelling is needed for a deeper understanding of the dynamics involved, the spin~$s$ models already exhibit unique characteristics that make them well-suited for comparison with observations. In order to test this effect, we plan to perform more realistic simulations with a more complex structure for the stellar system, including for example a disk component.

By varying the parameters of our simulations, we identified the scaling relations that characterise both the central core and the external envelope of the resulting haloes based on the initial solitons involved in the merger. These relations are expressed as a function of the number of solitons (the total mass), which may be non-integer. Using this information, it is possible to determine the free parameters of the ULDM halo profile for a fixed halo mass, including the characteristic core radius $r_c$, the transition parameter $r_{\epsilon}$, the envelope radius $r_s$ and maximum density $\rho_c$. This particular approach serves to construct equivalent configurations for the three models in terms of mass or average density, which is valuable for comparative analysis.

These parametric relations can be used to characterise profiles at different simulation stages, helping reduce computational costs. For example, running cosmological simulations up to a high redshift, such as $z > 3$, can be computationally efficient. From there, the profile parameters can be determined by extrapolation, allowing further analysis without requiring extensive simulation time. Moreover, thanks to the scaling relations we identified, we constructed equivalent host haloes for each spin~$s$ model with the same core size finding that the tidal disruption time is longer for \spin{0}. However, when fixing the total mass of the halo, the satellite survives longer for the \spin{2} model. We remark that the scaling relations and the results we derive from them hold within the context of our simplified approach, in which all initial solitons are identical. This setup is nonetheless useful to directly trace the impact on the dark matter profile components.




In summary, \spin{1} and \spin{2} ULDM models can help resolve some of the problems that have been discussed for \spin{0} ULDM simulations. The first concerns the cores of haloes observed in some galactic systems. While all three models form a core, in the \spin{0} case the core has higher central densities.~\citet{BaryonDriven} demonstrated that including baryons leads to cuspy scalar dark matter profiles, thereby reintroducing the tension with observational data that ULDM was thought to cure. This problem can be relaxed if the resultant haloes have lower central densities, as seen in the \spin{1} and \spin{2} models. The second problem is related to the discrepancy between the predicted velocity dispersion in \spin{0} models and observations. Higher-spin models could help resolve this tension by predicting lower values for this quantity. Finally, in the case of satellite systems, spin ULDM models may predict longer orbital decay times, depending on the specific characteristics of the host halo, which can be contrasted with observations of dwarf spheroidal galaxies.

\section*{Acknowledgements}

We wish to thank our HPC staff Josef Dvořáček for technical support throughout this project and Mustafa Amin for valuable correspondence. The research leading to these results has received support from the European Structural and Investment Funds and the Czech Ministry of Education, Youth and Sports (project No. FORTE---CZ.02.01.01/00/22\_008/0004632). FU and CS acknowledge support from MEYS through the INTER-EXCELLENCE II, INTER-COST grant LUC23115. CS acknowledges support from the Royal Society Wolfson Visiting Fellowship ``Testing the properties of dark matter with new statistical tools and cosmological data''. This article is based upon work from the COST Action COSMIC WISPers CA21106, supported by COST (European Cooperation in Science and Technology).

\section*{Data Availability}
The evolution of the energies of the 24 simulations for each spin, the spherically averaged density profiles (as shown in Fig.~\ref{fig:DMPRofMM}) and the scaling relations discussed in sec.~\ref{sec: URSDP} are available in Zenodo, at https://doi.org/10.5281/zenodo.14791353


\bibliographystyle{mnras}
\bibliography{references} 

@article{Niemeyer-review,
  title={Small-scale structure of fuzzy and axion-like dark matter},
  author={Niemeyer, Jens C},
  journal={Progress in Particle and Nuclear Physics},
  volume={113},
  pages={103787},
  year={2020},
  publisher={Elsevier}
}

@article{Feng_2010,
   title={Dark Matter Candidates from Particle Physics and Methods of Detection},
   volume={48},
   ISSN={1545-4282},
   url={http://dx.doi.org/10.1146/annurev-astro-082708-101659},
   DOI={10.1146/annurev-astro-082708-101659},
   number={1},
   journal={Annual Review of Astronomy and Astrophysics},
   publisher={Annual Reviews},
   author={Feng, Jonathan L.},
   year={2010},
   month=aug, pages={495–545} }

@article{Bertone_2018,
   title={A new era in the search for dark matter},
   volume={562},
   ISSN={1476-4687},
   url={http://dx.doi.org/10.1038/s41586-018-0542-z},
   DOI={10.1038/s41586-018-0542-z},
   number={7725},
   journal={Nature},
   publisher={Springer Science and Business Media LLC},
   author={Bertone, Gianfranco and Tait, Tim M. P.},
   year={2018},
   month=oct, pages={51–56} }

@article{Ferreira_2021,
   title={Ultra-light dark matter},
   volume={29},
   ISSN={1432-0754},
   url={http://dx.doi.org/10.1007/s00159-021-00135-6},
   DOI={10.1007/s00159-021-00135-6},
   number={1},
   journal={The Astronomy and Astrophysics Review},
   publisher={Springer Science and Business Media LLC},
   author={Ferreira, Elisa G. M.},
   year={2021},
   month=sep }

@article{Jain_2022,
   title={Polarized solitons in higher-spin wave dark matter},
   volume={105},
   ISSN={2470-0029},
   url={http://dx.doi.org/10.1103/PhysRevD.105.056019},
   DOI={10.1103/physrevd.105.056019},
   number={5},
   journal={Physical Review D},
   publisher={American Physical Society (APS)},
   author={Jain, Mudit and Amin, Mustafa A.},
   year={2022},
   month=mar }

@article{Alexander_2021,
   title={Higher spin dark matter},
   volume={819},
   ISSN={0370-2693},
   url={http://dx.doi.org/10.1016/j.physletb.2021.136436},
   DOI={10.1016/j.physletb.2021.136436},
   journal={Physics Letters B},
   publisher={Elsevier BV},
   author={Alexander, Stephon and Jenks, Leah and McDonough, Evan},
   year={2021},
   month=aug, pages={136436} }

@article{Hu-2000,
  title = {Fuzzy Cold Dark Matter: The Wave Properties of Ultralight Particles},
  author = {Hu, Wayne and Barkana, Rennan and Gruzinov, Andrei},
  journal = {Phys. Rev. Lett.},
  volume = {85},
  issue = {6},
  pages = {1158--1161},
  numpages = {0},
  year = {2000},
  month = {Aug},
  publisher = {American Physical Society},
  doi = {10.1103/PhysRevLett.85.1158},
  url = {https://link.aps.org/doi/10.1103/PhysRevLett.85.1158}
}

@article{Matos:1999et,
    author = "Matos, Tonatiuh and Guzman, Francisco Siddhartha and Urena-Lopez, L. Arturo",
    title = "{Scalar field as dark matter in the universe}",
    eprint = "astro-ph/9908152",
    archivePrefix = "arXiv",
    doi = "10.1088/0264-9381/17/7/309",
    journal = "Class. Quant. Grav.",
    volume = "17",
    pages = "1707--1712",
    year = "2000"
}

@article{TULIN20181,
title = {Dark matter self-interactions and small scale structure},
journal = {Physics Reports},
volume = {730},
pages = {1-57},
year = {2018},
note = {Dark matter self-interactions and small scale structure},
issn = {0370-1573},
doi = {https://doi.org/10.1016/j.physrep.2017.11.004},
url = {https://www.sciencedirect.com/science/article/pii/S0370157317304039},
author = {Sean Tulin and Hai-Bo Yu},
}

@article{Amin_2022,
   title={Small-scale structure in vector dark matter},
   volume={2022},
   ISSN={1475-7516},
   url={http://dx.doi.org/10.1088/1475-7516/2022/08/014},
   DOI={10.1088/1475-7516/2022/08/014},
   number={08},
   journal={Journal of Cosmology and Astroparticle Physics},
   publisher={IOP Publishing},
   author={Amin, Mustafa A. and Jain, Mudit and Karur, Rohith and Mocz, Philip},
   year={2022},
   month=aug, pages={014} }

@article{Adshead_2021,
   title={Self-gravitating vector dark matter},
   volume={103},
   ISSN={2470-0029},
   url={http://dx.doi.org/10.1103/PhysRevD.103.103501},
   DOI={10.1103/physrevd.103.103501},
   number={10},
   journal={Physical Review D},
   publisher={American Physical Society (APS)},
   author={Adshead, Peter and Lozanov, Kaloian D.},
   year={2021},
   month=may }

@article{Gorghetto_2022,
   title={Dark photon stars: formation and role as dark matter substructure},
   volume={2022},
   ISSN={1475-7516},
   url={http://dx.doi.org/10.1088/1475-7516/2022/08/018},
   DOI={10.1088/1475-7516/2022/08/018},
   number={08},
   journal={Journal of Cosmology and Astroparticle Physics},
   publisher={IOP Publishing},
   author={Gorghetto, Marco and Hardy, Edward and March-Russell, John and Song, Ningqiang and West, Stephen M.},
   year={2022},
   month=aug, pages={018} }

@article{Jain:2023ojg,
    author = "Jain, Mudit and Amin, Mustafa A. and Thomas, Jonathan and Wanichwecharungruang, Wisha",
    title = "{Kinetic relaxation and Bose-star formation in multicomponent dark matter}",
    eprint = "2304.01985",
    archivePrefix = "arXiv",
    primaryClass = "astro-ph.CO",
    doi = "10.1103/PhysRevD.108.043535",
    journal = "Phys. Rev. D",
    volume = "108",
    number = "4",
    pages = "043535",
    year = "2023"
}

@article{chen2023gravitational,
  title = {Gravitational Bose-Einstein condensation of vector or hidden photon dark matter},
  author = {Chen, Jiajun and Du, Xiaolong and Zhou, Mingzhen and Benson, Andrew and Marsh, David J. E.},
  journal = {Phys. Rev. D},
  volume = {108},
  issue = {8},
  pages = {083021},
  numpages = {10},
  year = {2023},
  month = {Oct},
  publisher = {American Physical Society},
  doi = {10.1103/PhysRevD.108.083021},
  url = {https://link.aps.org/doi/10.1103/PhysRevD.108.083021}
}

@article{Galic,
    author = {Yurin, Denis and Springel, Volker},
    title = "{An iterative method for the construction of N-body galaxy models in collisionless equilibrium}",
    journal = {Monthly Notices of the Royal Astronomical Society},
    volume = {444},
    number = {1},
    pages = {62-79},
    year = {2014},
    month = {08},
    issn = {0035-8711},
    doi = {10.1093/mnras/stu1421},
    url = {https://doi.org/10.1093/mnras/stu1421},
    eprint = {https://academic.oup.com/mnras/article-pdf/444/1/62/18503149/stu1421.pdf},
}

@article{PyUltraLightSI,
  title = {Modifying PyUltraLight to model scalar dark matter with self-interactions},
  author = {Glennon, Noah and Prescod-Weinstein, Chanda},
  journal = {Phys. Rev. D},
  volume = {104},
  issue = {8},
  pages = {083532},
  numpages = {14},
  year = {2021},
  month = {Oct},
  publisher = {American Physical Society},
  doi = {10.1103/PhysRevD.104.083532},
  url = {https://link.aps.org/doi/10.1103/PhysRevD.104.083532}
}

@article{Gosenca:2023yjc,
    author = "Gosenca, Mateja and Eberhardt, Andrew and Wang, Yourong and Eggemeier, Benedikt and Kendall, Emily and Zagorac, J. Luna and Easther, Richard",
    title = "{Multifield ultralight dark matter}",
    eprint = "2301.07114",
    archivePrefix = "arXiv",
    primaryClass = "astro-ph.CO",
    doi = "10.1103/PhysRevD.107.083014",
    journal = "Phys. Rev. D",
    volume = "107",
    number = "8",
    pages = "083014",
    year = "2023"
}

@article{Edwards_2018,
doi = {10.1088/1475-7516/2018/10/027},
url = {https://dx.doi.org/10.1088/1475-7516/2018/10/027},
year = {2018},
month = {oct},
publisher = {},
volume = {2018},
number = {10},
pages = {027},
author = {Faber Edwards and Emily Kendall and Shaun Hotchkiss and Richard Easther},
title = {PyUltraLight: a pseudo-spectral solver for ultralight dark matter dynamics},
journal = {Journal of Cosmology and Astroparticle Physics}
}

@article{volkerLarge,
    author = {May, Simon and Springel, Volker},
    title = "{Structure formation in large-volume cosmological simulations of fuzzy dark matter: impact of the non-linear dynamics}",
    journal = {Monthly Notices of the Royal Astronomical Society},
    volume = {506},
    number = {2},
    pages = {2603-2618},
    year = {2021},
    month = {06},
    issn = {0035-8711},
    doi = {10.1093/mnras/stab1764},
    url = {https://doi.org/10.1093/mnras/stab1764},
    eprint = {https://academic.oup.com/mnras/article-pdf/506/2/2603/39207201/stab1764.pdf},
}

@article{Guzman-2004,
  title = {Evolution of the Schr\"odinger-Newton system for a self-gravitating scalar field},
  author = {Guzm\'an, F. Siddhartha and Ure\~na-L\'opez, L. Arturo},
  journal = {Phys. Rev. D},
  volume = {69},
  issue = {12},
  pages = {124033},
  numpages = {18},
  year = {2004},
  month = {Jun},
  publisher = {American Physical Society},
  doi = {10.1103/PhysRevD.69.124033},
  url = {https://link.aps.org/doi/10.1103/PhysRevD.69.124033}
}

@book{numrecip,
author = {Press, William H. and Teukolsky, Saul A. and Vetterling, William T. and Flannery, Brian P.},
title = {Numerical Recipes 3rd Edition: The Art of Scientific Computing},
year = {2007},
isbn = {0521880688},
publisher = {Cambridge University Press},
address = {USA},
edition = {3}
}

@article{schive2014cosmic,
  title={Cosmic structure as the quantum interference of a coherent dark wave},
  author={Schive, Hsi-Yu and Chiueh, Tzihong and Broadhurst, Tom},
  journal={Nature Physics},
  volume={10},
  number={7},
  pages={496--499},
  year={2014},
  publisher={Nature Publishing Group UK London}
}

@article{Schive:2014hza,
    author = "Schive, Hsi-Yu and Liao, Ming-Hsuan and Woo, Tak-Pong and Wong, Shing-Kwong and Chiueh, Tzihong and Broadhurst, Tom and Hwang, W. -Y. Pauchy",
    title = "{Understanding the Core-Halo Relation of Quantum Wave Dark Matter from 3D Simulations}",
    eprint = "1407.7762",
    archivePrefix = "arXiv",
    primaryClass = "astro-ph.GA",
    doi = "10.1103/PhysRevLett.113.261302",
    journal = "Phys. Rev. Lett.",
    volume = "113",
    number = "26",
    pages = "261302",
    year = "2014"
}

@article{TidalSelf,
  title = {Tidal disruption of solitons in self-interacting ultralight axion dark matter},
  author = {Glennon, Noah and Nadler, Ethan O. and Musoke, Nathan and Banerjee, Arka and Prescod-Weinstein, Chanda and Wechsler, Risa H.},
  journal = {Phys. Rev. D},
  volume = {105},
  issue = {12},
  pages = {123540},
  numpages = {15},
  year = {2022},
  month = {Jun},
  publisher = {American Physical Society},
  doi = {10.1103/PhysRevD.105.123540},
  url = {https://link.aps.org/doi/10.1103/PhysRevD.105.123540}
}

@article{du2018tidal,
  title={Tidal disruption of fuzzy dark matter subhalo cores},
  author={Du, Xiaolong and Schwabe, Bodo and Niemeyer, Jens C and B{\"u}rger, David},
  journal={Physical Review D},
  volume={97},
  number={6},
  pages={063507},
  year={2018},
  publisher={APS}
}

@article{hui2017ultralight,
  title = {Ultralight scalars as cosmological dark matter},
  author = {Hui, Lam and Ostriker, Jeremiah P. and Tremaine, Scott and Witten, Edward},
  journal = {Phys. Rev. D},
  volume = {95},
  issue = {4},
  pages = {043541},
  numpages = {32},
  year = {2017},
  month = {Feb},
  publisher = {American Physical Society},
  doi = {10.1103/PhysRevD.95.043541},
  url = {https://link.aps.org/doi/10.1103/PhysRevD.95.043541}
}

@article{van2018disruption,
  title={Disruption of dark matter substructure: fact or fiction?},
  author={Van den Bosch, Frank C and Ogiya, Go and Hahn, Oliver and Burkert, Andreas},
  journal={Monthly Notices of the Royal Astronomical Society},
  volume={474},
  number={3},
  pages={3043--3066},
  year={2018},
  publisher={Oxford University Press}
}

@article{errani2021asymptotic,
  title={The asymptotic tidal remnants of cold dark matter subhaloes},
  author={Errani, Rapha{\"e}l and Navarro, Julio F},
  journal={Monthly Notices of the Royal Astronomical Society},
  volume={505},
  number={1},
  pages={18--32},
  year={2021},
  publisher={Oxford University Press}
}

@article{Courant1928,
  author    = {R. Courant and K. Friedrichs and H. Lewy},
  title     = {Über die partiellen Differenzengleichungen der mathematischen Physik},
  journal   = {Mathematische Annalen},
  year      = {1928},
  volume    = {100},
  number    = {1},
  pages     = {32--74},
  doi       = {10.1007/BF01448839},
  url       = {https://doi.org/10.1007/BF01448839},
}

@article{guzman2003newtonian,
  title={Newtonian collapse of scalar field dark matter},
  author={Guzm{\'a}n, F Siddhartha and Ure{\~n}a-L{\'o}pez, L Arturo},
  journal={Physical Review D},
  volume={68},
  number={2},
  pages={024023},
  year={2003},
  publisher={APS}
}

@article{BaryonDriven,
  title = {Baryon-driven growth of solitonic cores in fuzzy dark matter halos},
  author = {Veltmaat, Jan and Schwabe, Bodo and Niemeyer, Jens C.},
  journal = {Phys. Rev. D},
  volume = {101},
  issue = {8},
  pages = {083518},
  numpages = {7},
  year = {2020},
  month = {Apr},
  publisher = {American Physical Society},
  doi = {10.1103/PhysRevD.101.083518},
  url = {https://link.aps.org/doi/10.1103/PhysRevD.101.083518}
}

@article{smallscaleprob1,
   author = "Bullock, James S. and Boylan-Kolchin, Michael",
   title = "Small-Scale Challenges to the ΛCDM Paradigm", 
   journal= "Annual Review of Astronomy and Astrophysics",
   year = "2017",
   volume = "55",
   number = "Volume 55, 2017",
   pages = "343-387",
   doi = "https://doi.org/10.1146/annurev-astro-091916-055313",
   url = "https://www.annualreviews.org/content/journals/10.1146/annurev-astro-091916-055313",
   publisher = "Annual Reviews",
   issn = "1545-4282",
   type = "Journal Article",
  }

@Article{smallscaleprob2,
AUTHOR = {Del Popolo, Antonino and Le Delliou, Morgan},
TITLE = {Small Scale Problems of the ΛCDM Model: A Short Review},
JOURNAL = {Galaxies},
VOLUME = {5},
YEAR = {2017},
NUMBER = {1},
ARTICLE-NUMBER = {17},
URL = {https://www.mdpi.com/2075-4434/5/1/17},
ISSN = {2075-4434},
DOI = {10.3390/galaxies5010017}
}

@article{church2019heating,
  title={Heating of Milky Way disc stars by dark matter fluctuations in cold dark matter and fuzzy dark matter paradigms},
  author={Church, Benjamin V and Mocz, Philip and Ostriker, Jeremiah P},
  journal={Monthly Notices of the Royal Astronomical Society},
  volume={485},
  number={2},
  pages={2861--2876},
  year={2019},
  publisher={Oxford University Press}
}

@article{kawai2022analytic,
  title={An analytic model for the subgalactic matter power spectrum in fuzzy dark matter halos},
  author={Kawai, Hiroki and Oguri, Masamune and Amruth, Alfred and Broadhurst, Tom and Lim, Jeremy},
  journal={The Astrophysical Journal},
  volume={925},
  number={1},
  pages={61},
  year={2022},
  publisher={IOP Publishing}
}

@article{chowdhury2023dynamical,
  title={On the dynamical heating of dwarf galaxies in a fuzzy dark matter halo},
  author={Chowdhury, Dhruba Dutta and van den Bosch, Frank C and van Dokkum, Pieter and Robles, Victor H and Schive, Hsi-Yu and Chiueh, Tzihong},
  journal={The Astrophysical Journal},
  volume={949},
  number={2},
  pages={68},
  year={2023},
  publisher={IOP Publishing}
}

@article{dalal2022excluding,
  title={Excluding fuzzy dark matter with sizes and stellar kinematics of ultrafaint dwarf galaxies},
  author={Dalal, Neal and Kravtsov, Andrey},
  journal={Physical Review D},
  volume={106},
  number={6},
  pages={063517},
  year={2022},
  publisher={APS}
}

@article{Foreman-Mackey:2012any,
    author = "Foreman-Mackey, Daniel and Hogg, David W. and Lang, Dustin and Goodman, Jonathan",
    title = "{emcee: The MCMC Hammer}",
    eprint = "1202.3665",
    archivePrefix = "arXiv",
    primaryClass = "astro-ph.IM",
    doi = "10.1086/670067",
    journal = "Publ. Astron. Soc. Pac.",
    volume = "125",
    pages = "306--312",
    year = "2013"
}

@article{hunter2012little,
  title={Little things},
  author={Hunter, Deidre A and Ficut-Vicas, Dana and Ashley, Trisha and Brinks, Elias and Cigan, Phil and Elmegreen, Bruce G and Heesen, Volker and Herrmann, Kimberly A and Johnson, Megan and Oh, Se-Heon and others},
  journal={The Astronomical Journal},
  volume={144},
  number={5},
  pages={134},
  year={2012},
  publisher={IOP Publishing}
}

@ARTICLE{1990ApJHernquist,
       author = {{Hernquist}, Lars},
        title = "{An Analytical Model for Spherical Galaxies and Bulges}",
      journal = {\apj},
         year = 1990,
        month = jun,
       volume = {356},
        pages = {359},
          doi = {10.1086/168845},
       adsurl = {https://ui.adsabs.harvard.edu/abs/1990ApJ...356..359H}
}

@misc{amaral2024vectorwavedarkmatter,
      title={Vector Wave Dark Matter and Terrestrial Quantum Sensors}, 
      author={Dorian W. P. Amaral and Mudit Jain and Mustafa A. Amin and Christopher Tunnell},
      year={2024},
      eprint={2403.02381},
      archivePrefix={arXiv},
      primaryClass={hep-ph},
      url={https://arxiv.org/abs/2403.02381}, 
}

@article{PhysRevD.111.103520,
  title = {Cosmological perturbations with ultralight vector dark matter fields: Numerical implementation in class},
  author = {Ferreira Chase, Tomas and Leizerovich, Mat\'{\i}as and Lopez Nacir, Diana and Landau, Susana},
  journal = {Phys. Rev. D},
  volume = {111},
  issue = {10},
  pages = {103520},
  numpages = {19},
  year = {2025},
  month = {May},
  publisher = {American Physical Society},
  doi = {10.1103/PhysRevD.111.103520},
  url = {https://link.aps.org/doi/10.1103/PhysRevD.111.103520}
}




\appendix

\section{Ground state soliton}\label{a: GS-soliton}

For \spin{0}  particles, the system of equations \eqref{eq:multipleSP} is reduced to the well-known Schr\"odinger-Poisson system
\begin{align}\label{eq: SP-notime}
    \begin{aligned}
    i\hbar\frac{\partial}{\partial t}\psi&=-\frac{\hbar^2}{2m_s}\nabla^2\psi+m_s\Phi\psi\\
    \nabla^2\Phi&=4\pi G\abs{\psi}^2.
\end{aligned}
\end{align}
In spherical coordinates, the Laplacian can be expressed as 
\[
\nabla^2 f = \left(\frac{\partial^2}{\partial r^2}+\frac{2}{r}\frac{\partial}{\partial r}\right)f+\frac{1}{r^2\sin\theta}\frac{\partial}{\partial\theta}\left(\sin\theta\frac{\partial}{\partial\theta}\right)f+\frac{1}{r^2\sin^2\theta}\frac{\partial^2}{\partial\varphi^2}f,
\]
where $\theta$ and $\varphi$ are the polar and azimuthal angles. After assuming spherical symmetry, we can drop the angular dependency. Then, the Schr\"odinger-Poisson system takes the form
\begin{eqnarray}
    \frac{\hbar^2}{2m_s}\frac{\partial^2}{\partial r^2}(r\psi_{\text{sol}})&=&r\psi_{\text{sol}}(m_s\Phi-\mu c^2),\\
     \frac{\partial^2}{\partial r^2}(r\Phi)&=&4\pi G r\psi^2_{\text{sol}}.
\end{eqnarray}

\begin{figure}
    \centering
    \includegraphics[width=0.4\textwidth]{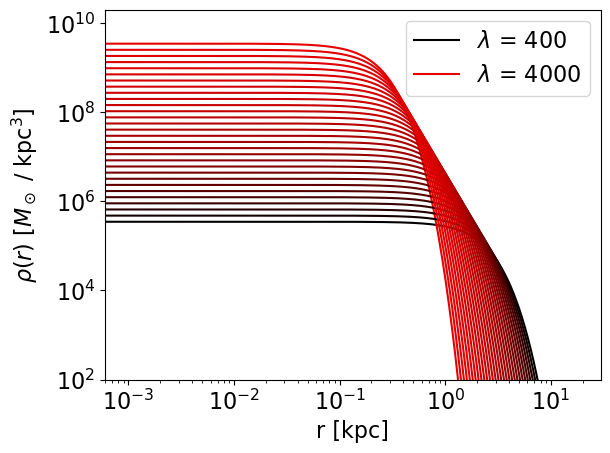}
    \caption{Ground state solution for the time-independent SP system considering $m=2.5\times 10^{-22}$ eV. The different colours show different values for the scaling parameter $\lambda$ to highlight the differences in the central and outer regions described in equation \eqref{eq:scaling}. In this paper we fixed $\lambda =1000$ for all the solitonic configurations used.}
    \label{fig:sol}
\end{figure}

It is convenient to rewrite the system of equations using the following transformations $\hat{\psi}_{\text{sol}}=\displaystyle\frac{\sqrt{G}}{\tilde{\hbar}}\psi_{\text{sol}}$ and $\hat{\Phi}=\displaystyle\frac{\Phi}{\tilde{\hbar}^2}$. Then, we have
\begin{eqnarray}\label{eq: eigensystem}
    \frac{1}{2}\frac{\partial^2}{\partial r^2}(r\hat{\psi}_{\text{sol}})&=&r\hat{\psi}_{\text{sol}}(\hat{\Phi}-\hat{\mu}),\\
    \frac{\partial^2}{\partial r^2}(r\hat{\Phi})&=&4\pi r \hat{\psi}_{\text{sol}},
\end{eqnarray}
where $\tilde{\hbar}=\displaystyle\frac{\hbar}{m_s}$ and $\hat{\mu}=\displaystyle\frac{m_s \mu c^2}{\hbar^2}$ is a constant which corresponds to the eigenvalue of the system \eqref{eq: eigensystem}. Since we are looking for equilibrium configurations, we consider the following conditions
\begin{align}\label{eq: scaling_relations}
    \begin{aligned}
        &\hat{\psi}_{\text{sol}}(r\rightarrow \infty) \rightarrow 0, & \quad &\hat{\Phi}(r\rightarrow\infty) = -\frac{GM}{\tilde{\hbar}^2r}, \\
        &\hat{\psi}(r\rightarrow 0) = 1, & \quad &\frac{\partial\hat{\Phi}}{\partial r}\bigg|_0 \rightarrow 0, \\
        &\frac{\partial\hat{\psi}}{\partial r}\bigg|_0 \rightarrow 0, & \quad &\frac{\partial\hat{\psi}}{\partial r}\bigg|_{r\rightarrow\infty} \rightarrow 0,
    \end{aligned}
\end{align}
with $M(r)=\int \rho dV=4\pi\int \rho(r)r^2dr$ is the enclosed mass at radius $r$ and $\rho/\rho_0=\abs{\psi}^2$. By setting these conditions, there are unique values of $\mu$ and $\Phi(0)$ for which the boundary conditions are fulfilled. Also, the SP system is invariant under the rescaling relations given by $\{M,m_\text{ULDM}\}\to  \{\lambda M, \beta m_\text{ULDM}\}$
\begin{equation}\label{eq:scaling}
    \left\{t, x , \psi, \rho\right\} \to \left\{\lambda^{-2}\beta^{-3}t, \lambda^{-1}\beta^{-2}x , \lambda^{2}\beta^{3}\psi,\lambda^{4} \beta^{6} \rho\right\}. 	
\end{equation}
We can find solutions for this system for a fixed value of the ULDM mass, as shown in Fig \ref{fig:sol}. See~\citet{Amin_2022,guzman2003newtonian} for further details.

\section{Performance of the code}\label{sec:Performance}

\begin{figure}
    \centering
    \includegraphics[width=0.4\textwidth]{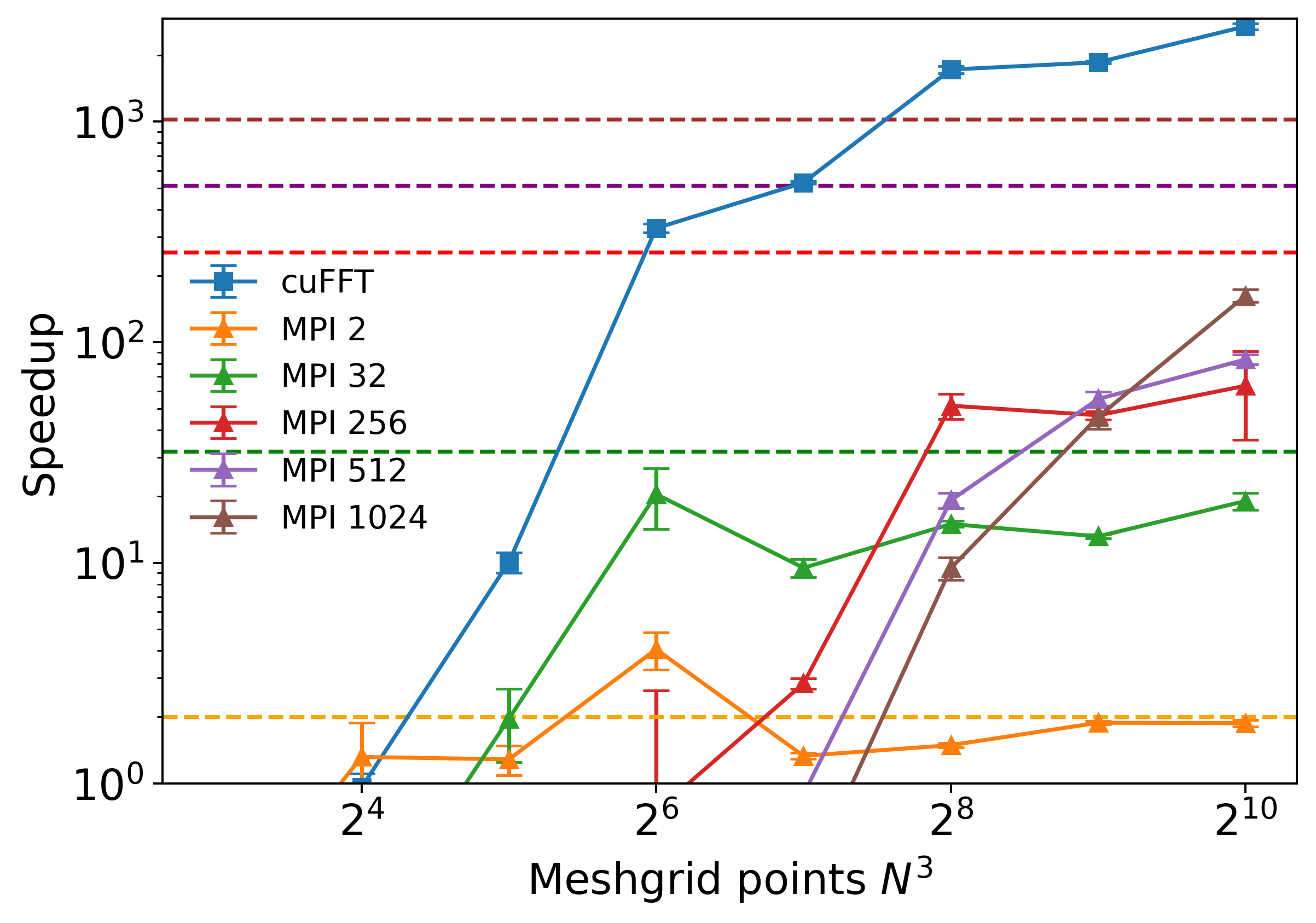}
    \caption{Comparison of the performance between the Fastest Fourier Transform in the West (FFTW) library using MPI with $2$, $32$, $256$ and $1024$ cores, and CUDA FFT (cuFFT) library on a single CPU. In all cases, the speed was computed relative to the performance of a single core. Horizontal dashed lines indicate the expected ideal performance for MPI.}
    \label{fig:Performance}
\end{figure}

To test the efficiency of the code, we have performed 1000 Fourier transformations using different mesh sizes to compare the execution time for two methods. The first one uses FFTW-MPI with different numbers of cores ($2$, $32$, $256$ and $1024$), while the second one uses cuFFT on a single Nvidia A100 GPU with 80GB of memory. Fig.~\ref{fig:Performance} shows the speedup of each method, defined as the ratio of its execution time to the execution time when using a single CPU. Theoretically, the speedup is expected to match the number of cores used. However, the process is inefficient due to communication between cores and memory allocation. The remarkable improvement the GPU provides becomes evident as the number of mesh grid points increases, significantly benefitting the type of simulations conducted in this work. The GPU performance is one order of magnitude larger than the MPI version.

One disadvantage of GPUs is their limited memory, whereas FFTW-MPI depends on RAM for memory allocation, which is usually larger than GPUs. We thus plan to use the MPI version of cuFFT (cuFFTMp) to increase allocation capacity in a future work.\footnote{\url{https://docs.nvidia.com/hpc-sdk/cufftmp/index.html}} Additionally, it is worth mentioning that we have parallelised the FFTW-MPI library for only one axis. Further improvements could involve parallelising in two dimensions, leading to better performance. There are publicly available tools, like 2decomp-fft, that can be used for this purpose.\footnote{\url{https://github.com/2decomp-fft/2decomp-fft}}

\section{Stability criterion for the GPP system (conservation of energy)}\label{sec: stability}

\begin{figure}
    \centering
    \includegraphics[width = 0.4\textwidth]{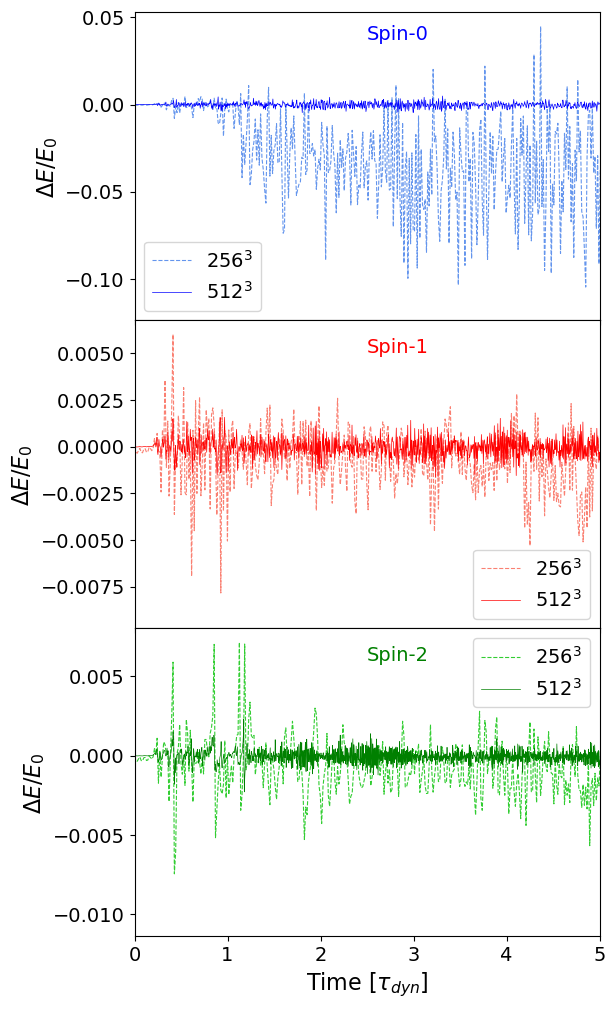}
    \caption{Energy variation as a function of time for \spin{0} (top, blue), \spin{1} (middle, green) and \spin{2} (bottom, red) considering a merger of $N_\text{sol}=20$, without loss of generality.}
    \label{fig:WEnergy}
\end{figure}

In Fig.~\ref{fig:WEnergy}, the evolution of the ratio $\Delta E/E_0$ over time is shown for 20 solitons, where $\Delta E=E(t)-E_0$ with $E_0$ the energy at $t=0$. Here $E = K+W$, where the kinetic and potential energies are defined in~\eqref{eq:Energy} and~\eqref{eq:Energy2}, respectively. This is presented for two different spatial resolutions,  $\Ngrid=256^3$ and $\Ngrid=512^3$, for \spin{0} (blue line), \spin{1} (green line) and \spin{2} (red line). The algorithm we apply, also known as a kick-drift-kick method, has second-order error $O(2)$ for the temporal step. Ideally, the energy should remain constant, with $\Delta E = 0$. However, some errors arise due to the finite approximation of the wave function. Our code demonstrates convergence as we increase the spatial resolution from $256^3$ (dashed lines) to $512^3$ (solid lines), with an exponential decrease in error. The \spin{0} case (blue) exhibits the most significant error propagation, primarily due to denser and narrower structures forming, which will require much more resolution than in the other models. In contrast, the \spin{2} case (red) shows lower error propagation, given that the central density of these haloes is smaller. As expected, all three spin~$s$ ULDM models converge to zero at the highest resolution. This works as a consistency test of the conservation of energy, with better convergence for lower resolutions in the case of higher spin, since constructive interference becomes less likely. Consequently, the \spin{0} model will require higher resolution than \spin{1} and \spin{2}, and the effects of varying resolutions will be more pronounced in the first case. The behaviour for a different number of solitons is similar. 

\begin{figure}
    \centering
    \includegraphics[width=0.8\linewidth]{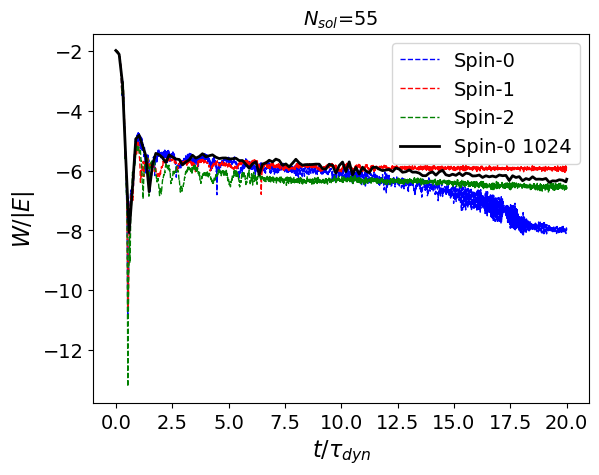}
    \caption{Evolution of the ratio $W/\abs{E}$ as a function of  $t/\tau_{\text{dyn}}$ for \spin{0}, \spin{1} and \spin{2} models with $N_\text{sol}=55$ (within the range where the stability criteria is not fulfilled for $\Ngrid=512^3$). The blue, red and green lines corresponds to \spin{0}, \spin{1} and \spin{2} with a resolution of $\Ngrid=512^3$. The black lines refers to the \spin{0} model with $\Ngrid=1024^3$.}
    \label{fig:resolution-test}
\end{figure}

Additionally, we tested the energy stability of the system by following the ratio $W/\abs{E}$ over time. We considered two different mesh resolutions $512^3$ and $1024^3$. Fig.~\ref{fig:resolution-test} shows the $W/\abs{E}$ ratio using $\Ngrid=512^3$ for the three models and $\Ngrid=1024^3$ for \spin{0} only. In all cases, $\Nsol=55$ was considered. We found that, while for \spin{0} there is a break in the curve due to the resolution, \spin{1} and \spin{2} show well-defined convergence behaviour. This becomes more evident for larger values of $\Nsol$. In fact, we can reproduce the same behaviour using both resolutions for \spin{0} if $\Nsol<30$. That is, below this threshold it is valid to use both spatial resolutions for the three models. As mentioned in section~\ref{sec: ics}, the \spin{0} model requires higher resolution because it has only one component for the density, which could lead to larger values near the resolution limit. This issue is not present in the other models, where the wavefunction can be split into more components. Therefore, using $\Ngrid = 512$ for \spin{1} and~2 is sufficient, and increasing the resolution would be unnecessarily computationally expensive.

\begin{figure}
    \centering
    \includegraphics[width=0.8\linewidth]{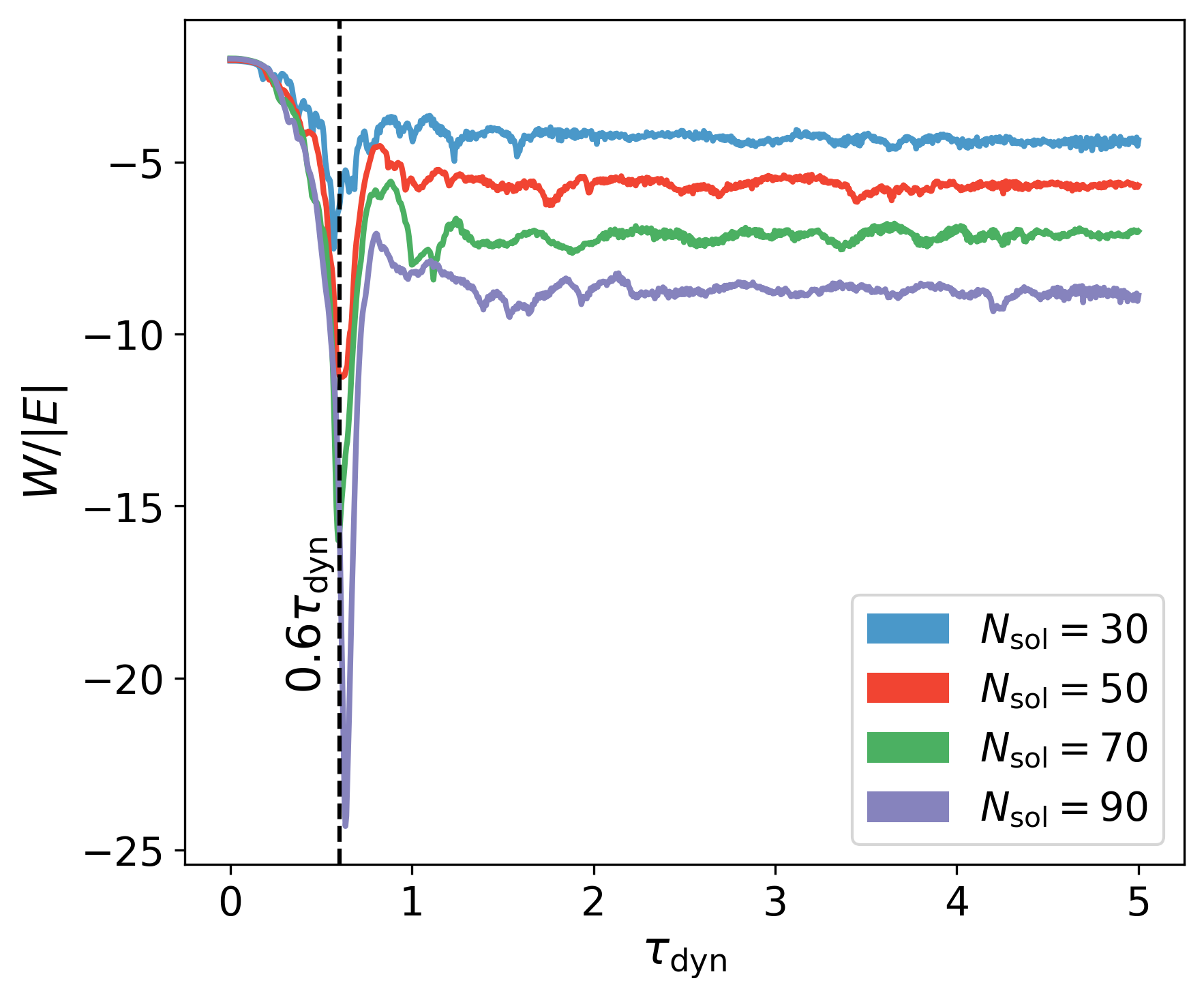}
    \caption{Evolution of the $W/\abs{E}$ ratio  for the spin-0 model as a function of $t/\tau_{\text{dyn}}$. We considered $N_{\text{sol}} = 30$, 50, 70, and 90 (ordered from top to bottom 
at large $\tau_\text{dyn}$) within a 100 kpc$^3$ box and a $512^3$ meshgrid. We observe that the merging begins around $0.6 \tau_{\text{dyn}}$ when most of the solitons merge to form more complex subhaloes, after which the system collapses and gradually relaxes. In this sense, the merging time is compared to the dynamical timescale $\tau_{\text{dyn}}$.}
    \label{fig:MergeTime}
\end{figure}

Finally, in~\ref{fig:MergeTime}, we show the evolution of the ratio $W/\abs{E}$ for four different simulations considering \spin{0} only to verify that $\tau_\text{dyn}$ is indeed appropriate to scale the time evolution of our simulations. The peak of the potential energy, which is correlated to the merging process, is shown as a dashed black line. We verify this for different numbers of solitons $N_\text{sol}$ and conclude that the system starts a relaxation process after $t\sim\tau_\text{dyn}$. 

\section{Multiple scalars}\label{sec: spinVsMulti}

The multi-field scenario is characterised by multiple scalar fields $\psi_i$ such that $\psi_i \propto c_i \psi$. The main difference compared to the spin-$s$ case of Eq.~\eqref{eq: components} is that,
 for the multi-field case, the mass distribution is global, leaving little room for local perturbations; larger perturbations away from the global value are statistically suppressed and are not expected to occur.
 In contrast, for spin-$s$ ULDM, each soliton has its own field configuration determined by its linear combination of polarisation states. This feature is particularly interesting in the spin-1 case, where one could choose to populate only the $\epsilon^{(0)}$ polarisation, and the evolution of that component would match that of spin-0, albeit projected only along the $z$-axis. On the other hand, in the polarisation basis of a spin-$s$ field each soliton can be only partially polarised, allowing for both gravitational and component-wise interactions, thereby leading to a distinct phenomenology.

\begin{figure}
    \centering
    \includegraphics[width=0.9\linewidth]{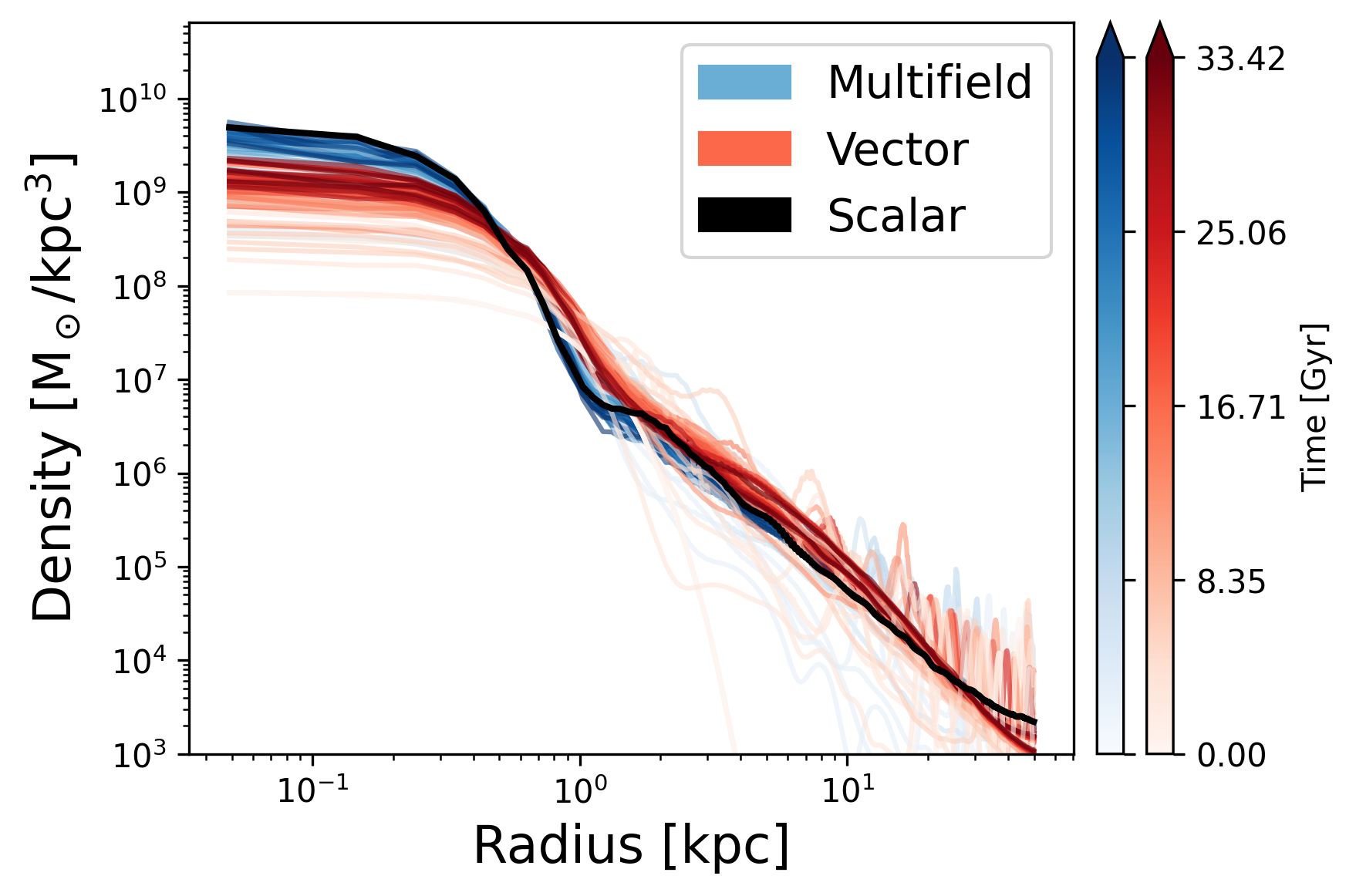}
    \caption{Density profiles after 5~$T_\text{dyn}$ following the merger of 15~initial solitons for 5~different initial conditions using a box of 100~kpc and $512^3$ mesh grid points. The curves (from top to bottom at small radii) represent, respectively; a scalar field, a three-component multiefield odel and a \spin{1} case. The gradation of tones (darker = later) shows the time evolution.}
    \label{fig: spin_vs_multi}
\end{figure}

In Fig.~\ref{fig: spin_vs_multi}, we show the differences between the multi-field and spin-1 models under identical initial conditions. From this plot, we can conclude that the multi-field model with equal boson mass yields a denser core, resembling the behaviour of the spin~$0$ case. This is expected, since the evolution of the three components is driven by equivalent fields with similar ground states. In contrast, the spin~$1$ scenario behaves differently: each soliton has its own matter distribution across the vector field components, resulting in distinct merging processes for each polarisation. In some cases, this leads to the early disruption of cores on a single component, highlighting the anisotropic dynamics of model.

\begin{figure}
    \centering
    \includegraphics[width=0.8\linewidth]{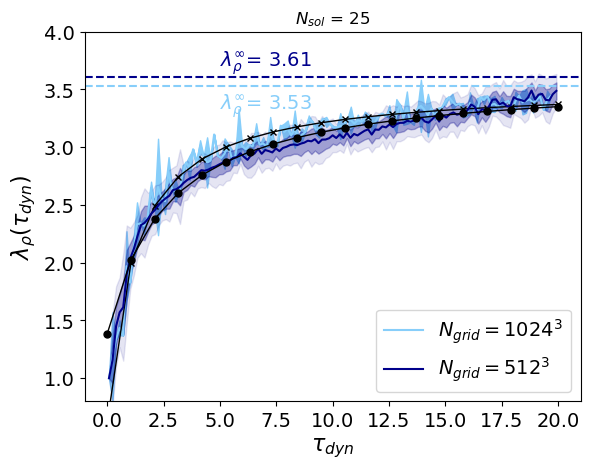}
    \caption{Evolution of $\lambda_{\rho}$ as a function of $\tau_{\text{dyn}}$ for \spin{0} considering two different resolutions for the simulation, $512^3$ and $1024^3$ for $\Nsol=25$ (within the range where the stability criteria of energy is fulfilled for both resolutions). The asymptotic convergence value at infinity is also shown for both cases. }
    \label{fig: resol_app}
\end{figure}

\section{Resolution tests for $\lambda_{\rho}$}\label{sec: lambda_conv}
As a complement of Appendix~\ref{sec: stability}, we compare the evolution of $\lambda_{\rho}$ as a function of $\tau_{\text{dyn}}$ using both resolutions for the \spin{0} model with $\Nsol=25$. This value of $\Nsol$ lies within the range where both resolutions, $\Ngrid=512^3$ and $\Ngrid=1024^3$, converge and satisfy the stability criterion. (see Appendix~\ref{sec: stability}) . In Fig.~\ref{fig: resol_app}, we present $\lambda_{\rho}$ for both cases, showing a similar behaviour for both scenarios. The corresponding values of $\lambda_{\rho}^{\infty}$ for each case are also included. Beyond $\Nsol>30$, there is no convergence for $\lambda_{\rho}$ with $\Ngrid=512^3$.

\section{Convergence rate for $\lambda^{\infty}_{\rho}$}\label{sec: lambda_infty}
\begin{figure}
    \includegraphics[width=0.9\linewidth]{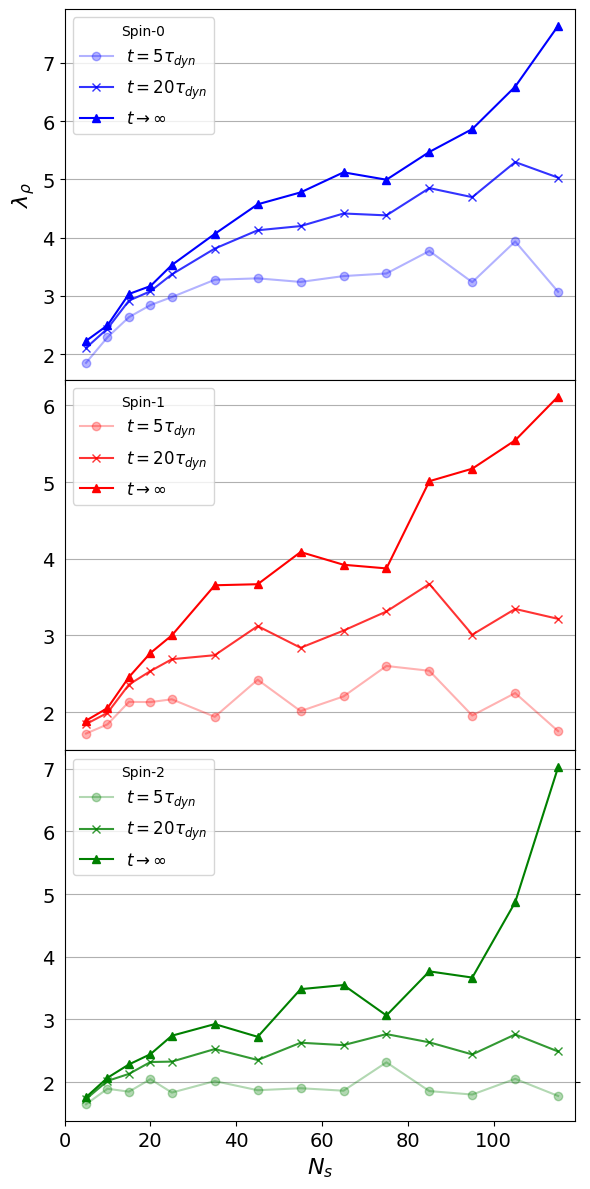}
    \caption{From top to bottom: Evolution of $\lambda_{\rho}$ for different $\Nsol$ for $t=\{5,20,\infty\}\tau_{\text{dyn}}$ for \spin{0} (left), \spin{1} (centre) and \spin{2} (right).}
    \label{fig:lambda_convergence}
\end{figure}

This section presents the evolution of $\lambda_{\rho}$ as a function of $\Nsol$ for different dynamical times. In Fig.~\ref{fig:lambda_convergence}, we observe the curves for this quantity at $t = \{5, 20, \infty\}\tau_{\text{dyn}}$, for the spin-0 (left), spin 1 (centre) and spin 2 (right) models. We found that the larger the dynamical time at which the densities from~\eqref{eq:lambdas} were computed, the greater the corresponding value of $\lambda_{\rho}$ for the same $\Nsol$. The saturation value is reached when $t\rightarrow \infty$. Due to computational limitations, in this work, we limit the simulations to $20\tau_{\text{dyn}}$ to perform the fits discussed in Sec.~\ref{sec: URSDP}. Additionally, we considered more than $5\tau_{\text{dyn}}$, since for the larger $N_\text{sol}$, the systems show that they satisfy the energy relaxation criteria, but some solitons still stay in orbit. On the other hand, we also observe that, for the same dynamical time, $\lambda_{\rho}$ is higher the lower the spin, due to the prominent cores for the spin 0 model, showing a very defined hierarchical evolution for this quantity. 

\bsp	
\label{lastpage}
\end{document}